# Adapting to unknown noise level in sparse deconvolution

Claire Boyer[1], Yohann De Castro[2], and Joseph Salmon[3]

[1]Institut de Mathmatiques de Toulouse, INSA Toulouse, 135 av. de Rangueil 31400 Toulouse, France
[2]Laboratoire de Mathématiques d'Orsay, Univ. Paris-Sud, CNRS, Université Paris-Saclay, 91405 Orsay, France
[3]LTCI, CNRS, Télécom ParisTech, Université Paris-Saclay, 75013, Paris, France

October 19, 2016

**Abstract**

In this paper, we study sparse spike deconvolution over the space of complex-valued measures when the input measure is a finite sum of Dirac masses. We introduce a modified version of the Beurling Lasso (BLasso), a semi-definite program that we refer to as the Concomitant Beurling Lasso (CBLasso). This new procedure estimates the target measure and the unknown noise level simultaneously. Contrary to previous estimators in the literature, theory holds for a tuning parameter that depends only on the sample size, so that it can be used for unknown noise level problems. Consistent noise level estimation is standardly proved. As for Radon measure estimation, theoretical guarantees match the previous state-of-the-art results in Super-Resolution regarding minimax prediction and localization. The proofs are based on a bound on the noise level given by a new tail estimate of the supremum of a stationary non-Gaussian process through the Rice method.

# 1 Introduction

## 1.1 Sparse deconvolution with unknown noise

### 1.1.1 Super-Resolution

Sparse deconvolution over the space of complex-valued Borel measures has recently attracted a lot of attention in the "Super-Resolution" community and its companion formulation "Line spectral estimation". In the Super-Resolution framework, one aims at recovering fine scale details of an image from few low frequency measurements, where ideally the observation is given by a low-pass filter. The novelty in this body of work relies on new theoretical guarantees of the $\ell_1$-minimization over the space of discrete measures in a grid-less manner. Some recent works on this topic (when the underlying dimension is one) can be found in [18, 12, 13, 38, 14, 2, 25, 9, 22] and references therein.

More precisely, pioneering works were proposed in [12] treating inverse problems on the space of Borel measures and in [13], where the Super-Resolution problem was investigated via Semi-Definite Programming and a groundbreaking construction of a "dual certificate". Exact recovery (in the noiseless case), minimax prediction and localization (in the noisy case) have been performed using the Beurling Lasso (BLasso) estimator [2, 38, 25, 37] which minimizes the total variation norm over complex-valued Borel measures. Noise robustness (as the noise level tends to zero) has been thoroughly investigated in [22]; the reader may also consult [23, 20, 24] for more details. Change point detection and grid-less spline decomposition are studied in [7, 19]. Several interesting extensions, such as deconvolution over spheres, have been also recently provided in [6, 9, 8]. For more general settings, we refer to the work of [30].

#### 1.1.2 Concomitant Beurling Lasso: adapting to the noise

Our proposed estimator is an adaptation to the Super-Resolution framework of a methodology first developed for sparse high dimensional regression. In the latter case, the joint estimation of the parameter and of the noise level has first been considered in [33, 1], though without any theory. It was based on concomitant estimation ideas that could be traced back to the work of Huber [29]. The formulation we consider in this work appeared first in [33, 1] with a statistical point of view, as well as in [40] with a game theory flavor. Note that interestingly, both approaches rely on the notion of robustness. An equivalent definition of this estimator was proposed and extensively studied independently in [5] under the name Square-root Lasso. The formulation we investigate is also closer to the one analyzed in [36] under the name Scaled-Lasso. Yet, we adopt the terminology of "Concomitant Beurling Lasso" in reference to the seminal paper [33]. Last but not least, our contribution borrows some ideas from the stimulating lecture notes [39].

Remark that an alternative formulation was investigated in [35] with a particular aim at Gaussian mixture models. The authors have proposed to analyze a different high dimensional regression variant that also leads to a jointly convex (w.r.t. both the parameter and the noise level) reformulation of a penalized log-likelihood estimator. It is to be noted that this estimator is also sometimes referred to Scaled-Lasso, creating possible ambiguities. In practice though, at least in high dimensional regression settings, this method seems to be outperformed by the concomitant formulation [32].

### 1.2 Model and contributions

#### 1.2.1 Model and notation

Denote $E := (\mathcal{C}(\mathbb{T}, \mathbb{C}), \|\cdot\|_\infty)$ the space of complex-valued continuous functions over the one dimensional torus $\mathbb{T}$ (obtained by identifying the endpoints on $[0,1]$) equipped with the $\ell_\infty$-norm and $E^* := (\mathcal{M}(\mathbb{T}, \mathbb{C}), \|\cdot\|_{\mathrm{TV}})$ its dual topological space. Namely, $E^*$ is the space of complex-valued Borel measures over the torus endowed with the total variation norm, defined by

$$\forall \mu \in E^*, \quad \|\mu\|_{\mathrm{TV}} := \sup_{\|f\|_\infty \le 1} \Re\left(\int_{\mathbb{T}} \bar{f} \mathrm{d}\mu\right) \ , \tag{1}$$

where $\Re(\cdot)$ denotes the real part and $\bar{f}$ the complex conjugate of a continuous function $f$. Our observation vector is $y \in \mathbb{C}^n$ (where $n = 2f_c + 1$) and our sampling scheme is modeled by the linear operator $\mathcal{F}_n$ that maps a Borel measure to its $n$ first Fourier coefficients as

$$\forall \mu \in E^* , \quad \mathcal{F}_n(\mu) := (c_k(\mu))_{|k|\leq f_c}, \quad \text{where} \quad c_k(\mu) := \int_{\mathbb{T}} \exp(-2\pi\imath kt)\mu(\mathrm{d}t) = \int_{\mathbb{T}} \overline{\varphi_k}\mathrm{d}\mu \ ,$$

and $\varphi_k(\cdot) = \exp(2\pi\imath k\cdot)$. The statistical model we consider is formulated as follows

$$y = \mathcal{F}_n(\mu^0) + \varepsilon \ , \tag{2}$$

with $\varepsilon$ is a complex valued centered Gaussian random variable defined by $\varepsilon \overset{d}{=} \varepsilon^{(1)} + \imath\varepsilon^{(2)}$ where the real part $\varepsilon^{(1)} = \Re(\varepsilon)$ and the imaginary part $\varepsilon^{(2)} = \Im(\varepsilon)$ are i.i.d. $\mathcal{N}_n(0, \sigma_0^2 \mathrm{Id}_n)$ random vectors with an unknown standard deviation $\sigma_0 > 0$, where $\mathrm{Id}_n$ is the identity matrix of size $n \times n$. Moreover, we assume that the target measure $\mu^0$ admits a sparse structure, namely it has finite support and can be written

$$\mu^0 = \sum_{j=1}^{s_0} a_j^0 \delta_{t_j^0} \ , \tag{3}$$

where $s_0 \geq 1$, $\delta_{t_j^0}$ is the Dirac measure at positions $t_j^0 \in \mathbb{T}$ and with amplitudes $a_j^0 \in \mathbb{C}$. We can now introduce our Concomitant Beurling Lasso (CBLasso) estimator, that jointly estimates the signal and the noise level as the solution of the convex program

$$(\hat{\mu}, \hat{\sigma}) \in \underset{(\mu,\sigma)\in E^* \times \mathbb{R}_{++}}{\arg\min} \ \frac{1}{2n\sigma}\|y - \mathcal{F}_n(\mu)\|_2^2 + \frac{\sigma}{2} + \lambda\|\mu\|_{\mathrm{TV}} \ , \tag{4}$$

where $\mathbb{R}_{++}$ denotes the set of positive real numbers and $\lambda > 0$ is a tuning parameter. This formulation, by using a suitable rescaling of the data fitting and adding a penalty on the noise level, leads to a jointly convex formulation that can be theoretically analyzed. The division by $\sigma$ is used for homogeneity reasons, while the $\sigma/2$ term helps avoiding degenerate solutions and plays a regularization role.

When the solution is reached for $\hat{\sigma} > 0$, one can check that our estimator satisfies the identity $\hat{\sigma} = \|y - \mathcal{F}_n(\hat{\mu})\|_2/\sqrt{n}$ and $\hat{\mu} \in \arg\min_{\mu\in E^*} \|y - \mathcal{F}_n(\mu)\|_2/\sqrt{n} + \lambda\hat{\sigma}\|\mu\|_{\mathrm{TV}}$, which is in our framework, the analogous version of the square-root formulation from [5] (while the one from (4) is inspired by [33, 36]).

**Remark 1.** *As defined in (4), the CBLasso estimator is ill-defined. Indeed, the set over which we optimize is not closed and the optimization problem may have no solution. We circumvent this difficulty by considering instead the Fenchel biconjugate of the objective function. The actual objective function accepts $\sigma \geq 0$ as soon as $y = \mathcal{F}_n(\mu)$. In the rest of the paper, we will write (4) instead of the minimization of the biconjugate as a slight abuse of notation (see also [32] for more details).*

This new estimator can be efficiently computed using Fenchel-Legendre duality and a semi-definite representation of non-negative trigonometric polynomials. The dual program estimates the coefficients of a non-constant trigonometric polynomial (that we refer to as "dual polynomial") and the support of the estimated measure $\hat{\mu}$ is included in the roots of the derivative of the dual polynomial, see Section 3.1 for further details.

### 1.2.2 Contributions

By tackling the simultaneous estimation of the noise level and the target measure, we revisit the state-of-the-art results in Super-Resolution theory. In particular, we show (Theorem 1) that the "near" minimax prediction (*i.e.,* "fast rate" of convergence) is achieved by our new CBLasso estimator. To prove this result, we have adapted the proof of [37] to our estimator and finely controlled the noise level dependency in their bounds. This latter task has been carried out thanks to the Rice method for a non-Gaussian process (Lemma 19) which provides new results in this context, whose interest could go behind the context of Super-Resolution. Though standardly proved as in [2, 25, 37], spike localization errors (Theorem 2) are amended by the Rice method as well. In particular, it allows us control the "no-overfitting" event as shown by Proposition 4. We would like to emphasize that our contribution provides the first result on simultaneous estimation of both the noise level and the the target measure in spike deconvolution. We have introduced a new estimator and theoretically demonstrated that it attains "near" minimax optimal prediction together with strong localization accuracy. On the numerical side, we show that (i) the root-finding search can still be adapted to our method; (ii) the constructed "dual polynomial" (see Eq. (14) and (15) for definition) is never constant (see Proposition 7) proving the applicability of our method. Experiments are conducted to illustrate the benefits of our noise and measure estimation procedure.

## 1.3 Notation

We denote by $[m]$ the set $\{1,\dots,m\}$ for any integer $m \in \mathbb{N}$ and by $\mathbb{S}^{m-1}$ the (real) unit sphere in $\mathbb{R}^m$. For any set $A$, its indicator function and its cardinality respectively reads $\mathbb{1}_A$ and $|A|$. We denote by $\overline{z}$ the complex conjugate of $z \in \mathbb{C}$, and by $\Re(z)$ (resp. $\Im(z)$) its real (resp. imaginary) part. For any bounded linear mapping $\mathcal{F}$, its adjoint operator is denoted by $\mathcal{F}^*$. The standard Hermitian norm on $\mathbb{C}^n$ is written $\|\cdot\|$, with $\langle\cdot,\cdot\rangle$ being the associated inner product, *i.e.,* $\langle z, z'\rangle = \Re(z^* z')$. If a measure $\mu \in E^*$ can be written $\mu = \sum_{j=1}^s a_j \delta_{t_j}$ , we say that it has a finite support and we denote it by $\mathrm{supp}(\mu) := \{t_1,\dots,t_s\} \subset \mathbb{T}$. The canonical distance between two points $t$ and $t'$ on the torus $\mathbb{T}$ is written $\mathrm{d}(t,t')$.

# 2 Main results

## 2.1 Standard assumptions

In the CBLasso analysis, the following standard assumptions will be useful, see [39] for instance. The first assumption governs the Signal-to-Noise Ratio (SNR) that can be defined as

$$\mathrm{SNR} := \frac{\|\mu^0\|_{\mathrm{TV}}}{\sqrt{\mathbb{E}[\|\varepsilon\|_2^2]/n}} = \frac{\|\mu^0\|_{\mathrm{TV}}}{\sqrt{2}\sigma_0} \ ,$$

measuring the strength of the true signal $\mu^0$ compared to the noise level $\sigma_0$.

**Assumption 1** (Sampling rate condition)**.** *The sampling rate condition holds if and only if*

$$\lambda \cdot \mathrm{SNR} \leq \frac{\sqrt{17}-4}{2} \simeq 0.0616 \ . \tag{5}$$

The main point of the article is to consider a noise-free tuning parameter $\lambda$ that depends only on the number $n$ of measurements. As standard results in the literature, we consider $\lambda \geq 2\sqrt{2\log n/n}$. In this regime, one may write the sampling rate condition as $n/\log n \geq C\,\mathrm{SNR}^2$ for some universal constant $C > 0$. Roughly speaking, Assumption 1 states that the number of measurements $n$ is at least $\mathrm{SNR}^2$.

Another important assumption is the "no-overfitting" condition assuming that the noise level estimator $\hat{\sigma}$ is positive. If it does not hold, then the observations are perfectly fitted with our estimator and the residuals vanish. This kind of situation could happen when the noise level is small. Hence, Assumption 1 requires an upper bound on the signal-to-noise ratio (SNR) that could seem counterintuitive. However it ensures to have enough noise compared to signal to estimate it. For obvious reasons, it is essential from both theoretical and practical point of views to assert this property. Observe that, throughout this paper, all our results are based on the event $\{\|\mathcal{F}_n^*(\varepsilon)\|_\infty/(\sqrt{n}\|\varepsilon\|_2) \leq R\} \cap \{\|\varepsilon\|_2/\sqrt{n} \geq \underline{\sigma}\}$ for suitable $R$ and $\underline{\sigma}$, and by Proposition 4 the "no-overfitting" condition holds with large probability (in particular, note that Assumption 1 implies Inequality (8) with $\eta = 1/2$ whenever $n \geq -8\log\alpha$).

In Super-Resolution, one often assumes that the target measure $\mu^0 \in E^*$ satisfies the classical separation condition, see [26] for a state-of-the-art result on the subject. This condition governs the existence of dual certificates, see Section C for further results. In particular, all our constructions assume a lower bound on the number of observed frequencies $f_c$. Based on [26], we assume that $f_c \geq 10^3$ throughout this paper leading to $c_0 = 1.26$ in Assumption 2 below. Note that one can lower the bound on the observed frequencies $f_c$ considering larger values of $c_0$, the interested reader may consult [26] on this topic.

**Assumption 2** (Separation condition)**.** *The true support* $\mathrm{supp}(\mu^0) = \{t_1^0, \ldots, t_{s_0}^0\}$ *is said to verify the separation condition if it satisfies the following property*

$$\forall i, j \in [s_0], \quad s.t. \quad i \neq j, \qquad \mathrm{d}(t_i^0, t_j^0) \geq \frac{c_0}{f_c} \ ,$$

*where* $c_0 = 1.26$ *and* $f_c \geq 10^3$.

## 2.2 Compatibility limits

In order to obtain oracle inequalities for the Lasso [39], the statistical community has proposed various sufficient conditions such that Restricted Isometry Property (RIP), Restricted Eigenvalue Condition (REC), or Compatibility Condition for instance. However, in the Super-Resolution setting, one can show that these classical assumptions do not hold. Indeed, since RIP implies REC which in turn implies the Compatibility Condition (see [39] for further details), we only show that the Compatibility Condition fails to hold. To do so, let us recall the definition of the compatibility constant, denoted by $\mathcal{C}(L, S)$ for a constant $L > 0$ and a given support $S$

$$\mathcal{C}(L,S) := \inf\left\{|S|\|\mathcal{F}_n(\nu)\|_2^2/n; \quad \nu \in E^*, \|\nu_S\|_{\mathrm{TV}} = 1,\ \|\nu_{S^c}\|_{\mathrm{TV}} \leq L\right\} \ ,$$

We say that the compatibility condition of parameter $(s, L)$ holds if $\inf_{|S|\leq s} \mathcal{C}(L, S) > 0$. Here $\nu$ can be viewed as the difference of the two measures $\hat{\mu}$ and $\mu^0$, both verifying the separation condition (Assumption 2). Note that $\nu$ does not necessarily verify this assumption: two measures verifying Assumption 2 could have their supports as close as possible. Assume also that $L \geq 1$. Consider the sequence $(\nu_\epsilon)_{\epsilon>0}$ defined by $\nu_\epsilon = \delta_\epsilon - \delta_{-\epsilon}$, in which the location $-\epsilon$ on $\mathbb{T}$ can be associated to $1-\epsilon$. Note that for this sequence, we have $c_k(\nu_\epsilon) = -2\imath\sin(2\pi k\epsilon)$ for all $k \in \mathbb{Z}$, therefore $\|\mathcal{F}_n(\nu_\epsilon)\|_2^2 = \sum_{k=-f_c}^{f_c} 4\sin^2(2\pi k\epsilon)$. Let $\epsilon > 0$ and choose $S = S_\epsilon = \{\epsilon\}$ meaning that $|S_\epsilon| = 1$ and ${S_\epsilon}^c = \mathbb{T} \setminus \{\epsilon\}$, which leads to $\|(\nu_\epsilon)_{S_\epsilon}\|_{\mathrm{TV}} = 1$ and $\|(\nu_\epsilon)_{{S_\epsilon}^c}\|_{\mathrm{TV}} = 1 \leq L$. Considering $\epsilon \to 0$ leads to the inequality $\inf_{|S|\leq 1} \mathcal{C}(L,S) \leq \liminf_{\epsilon\to 0} \frac{1}{n}\sum_{k=-f_c}^{f_c} 4\sin^2(2\pi k\epsilon) = 0$. Since one can show that for $\underline{S} \subset S$, $\mathcal{C}(L,S)/|S| \leq \mathcal{C}(L,\underline{S})/|\underline{S}|$ [39], we deduce that for $s \geq 1$, $\inf_{|S|\leq s} \mathcal{C}(L,S) = 0$, which implies that the Compatibility Condition does not hold for Super-Resolution, and neither do the RIP or REC. It turns out that our setting meets the curse of highly correlated designs since close Dirac masses (as aforementioned) share almost the same Fourier coefficients.

## 2.3 Prediction error

As in [37], we uncover that CBLasso achieves the minimax rate[1] in prediction up to a log factor. Up to several technicalities, our proof in Appendix A follows the same guidelines as in [37] though we use the Rice method to finely bound our non-Gaussian process, see Lemma 19.

**Theorem 1.** *Let $C > 2\sqrt{2}$. There exists numerical constants $\gamma, C' > 0$, that may depend on $C$, such that the following holds. Under Assumptions 1 and 2, the estimator $\hat{\mu}$ solution to Problem 4 with a choice $\lambda \geq C\sqrt{\log n/n}$ satisfies*

$$\frac{1}{n}\|\mathcal{F}_n(\hat{\mu} - \mu^0)\|_2^2 \leq C'\, s_0\, \lambda^2\, \sigma_0^2 \ ,$$

*with probability at least $1 - C'n^{-\gamma}$.*

Up to a log factor, this prediction error bound matches the "fast rate" of convergence, namely $\sigma_0^2 s_0/n$ (see [17] for instance), established in sparse regression.

## 2.4 Localization and amplitudes estimation

Following [14, 26], we define the set of "near" points as

$$\forall j \in [s_0], \quad N_j := \left\{ t \in \mathbb{T};\ \mathrm{d}(t, t_j^0) \leq \frac{c_1}{f_c} \right\} \ , \tag{6}$$

for some $0 < c_1 < c_0/2$ (where $c_0 = 1.26$ is given in Assumption 2) and the set of "far" points as

$$F := \mathbb{T} \setminus \bigcup_{j\in[s_0]} N_j \ . \tag{7}$$

**Theorem 2.** *Let $C > 2\sqrt{2}$. There exist numerical constants $\gamma, C' > 0$, that may depend on $C$, such that the following holds. Suppose that Assumptions 1 and 2 hold. The estimator $\hat{\mu}$, solution to Problem 4 with a choice $\lambda \geq C\sqrt{\log n/n}$, satisfies*

1. $\forall j \in [s_0], \quad \Big| a_j^0 - \sum_{\{k:\ \hat{t}_k \in N_j\}} \hat{a}_k \Big| \leq C'\sigma_0 s_0 \lambda \ ,$

2. $\forall j \in [s_0], \quad \sum_{\{k:\ \hat{t}_k \in N_j\}} |\hat{a}_k| \mathrm{d}^2(t_j^0, \hat{t}_k) \leq C'\sigma_0 s_0 \lambda / n^2 \ ,$

[1] In [37], the minimax rate is derived using the minimax rate of [15] established in high-dimension statistics.

*3.* $\sum_{\{k:\ \hat{t}_k \in F\}} |\hat{a}_k| \leq C'\sigma_0 s_0 \lambda$ ,

*with probability at least* $1 - C'n^{-\gamma}$*.*

Points 1. and 3. in Theorem 2 ensure that $\hat{\mu}$ will retrieve the mass of $\mu^0$ in the near regions of the support $\operatorname{supp}(\mu^0)$ and not in regions far away. Point 2. provides a control on the support identification of the procedure. A proof of this theorem can be found in Section B. In particular, we deduce the following result.

**Corollary 3.** *Under the assumptions of Theorem 2, for any* $t_j^0$ *in the support of* $\mu^0$ *such that* $a_j^0 > C'\sigma_0 s_0 \lambda$*, there exists an element* $\hat{t}_k$ *in the support of* $\hat{\mu}$ *such that*

$$\mathrm{d}(t_j^0, \hat{t}_k) \leq \sqrt{\frac{C'\sigma_0 s_0 \lambda}{|a_j^0| - C'\sigma_0 s_0 \lambda} \frac{1}{n}} \ ,$$

*with probability at least* $1 - C'n^{-\gamma}$*.*

## 2.5 Noise level estimation

The following noise level estimation result relies on standard results in sparse regression, see [39].

**Proposition 4.** *Let* $0 < \eta < 1$ *and* $0 < \alpha < 1$*. Let* $\lambda$ *be the tuning parameter of the CBLasso. Set* $\underline{\sigma} = \sqrt{2}\sigma_0 \left(1 - \sqrt{-2\log\alpha/n}\right)^{1/2}$ *and* $R = \sqrt{2\log(n/\alpha)/n}$*. If* $\lambda \geq (1-\eta)^{-1}R$ *and*

$$\lambda \frac{\|\mu^0\|_{\mathrm{TV}}}{\underline{\sigma}} \leq 2\left[\sqrt{1 + (\eta/2)^2} - 1\right] \ , \tag{8}$$

*then it holds*

$$\left|\frac{\sqrt{n}\hat{\sigma}}{\|\varepsilon\|_2} - 1\right| \leq \eta \ , \tag{9}$$

*with probability larger than* $1 - \alpha\left(\frac{2\sqrt{2}}{n} + \frac{2\sqrt{3}+3}{3}\right)$*.*

Note that Assumption 1 implies Inequality (8) with $\eta = 1/2$ whenever $n \geq -8\log\alpha$.

*Proof.* The proof is a direct application of Lemma 19 with $R = \sqrt{2}\sqrt{\log(n/\alpha)/n}$ that gives

$$\mathbb{P}\left\{\frac{\|\mathcal{F}_n^*(\varepsilon)\|_\infty}{\sqrt{n}\|\varepsilon\|_2} \geq R\right\} \leq \alpha\left(\frac{2\sqrt{2}}{n} + \frac{2}{\sqrt{3}}\right) \ .$$

Applying Lemma 24 with $x = -\log\alpha$ gives $\mathbb{P}(\|\varepsilon\|_2/\sqrt{n} \leq \underline{\sigma}) \leq \alpha$. A union bound on the event $\{\|\mathcal{F}_n^*(\varepsilon)\|_\infty/(\sqrt{n}\|\varepsilon\|_2) \leq R\} \cap \{\|\varepsilon\|_2/\sqrt{n} \geq \underline{\sigma}\}$ combined with Proposition 17 finishes the proof. □

# 3 Numerical aspects

## 3.1 Primal/Dual problems and Fermat conditions

We begin by presenting the Fenchel dual formulation of CBLasso in the next proposition.

**Proposition 5.** *Denoting* $\mathcal{D}_n = \left\{c \in \mathbb{C}^n; \quad \|\mathcal{F}_n^*(c)\|_\infty \leq 1, n\lambda^2 \|c\|^2 \leq 1\right\}$*, the dual formulation of the CBLasso reads*

$$\hat{c} \in \operatorname*{arg\,max}_{c \in \mathcal{D}_n} \lambda \langle y, c \rangle \quad . \tag{10}$$

*Then, we have the link-equation between primal and dual solutions*

$$y = n\hat{\lambda}\hat{c} + \mathcal{F}_n(\hat{\mu}) \quad . \tag{11}$$

*where we define* $\hat{\lambda} = \lambda\hat{\sigma}$*, as well as a link between the coefficient and the polynomial*

$$\mathcal{F}_n^*(\hat{c}) = \hat{p} \quad . \tag{12}$$

*The polynomial* $\hat{p}$ *is said to be the dual polynomial of Problem* (4)*.*

*Proof.* This proposition is proved in Appendix E.1. □

Using (11) and (12), we retrieve the KKT conditions, namely

$$\frac{1}{n}\mathcal{F}_n^*(y - \mathcal{F}_n(\hat{\mu})) = \hat{\lambda}\hat{p} \quad , \tag{13}$$

In particular, the dual polynomial satisfies the property of a TV-norm sub-gradient at the solution point $\hat{\mu}$, namely

$$\|\hat{p}\|_\infty \leq 1, \tag{14}$$

$$\Re\left(\int_{\mathbb{T}} \overline{\hat{p}}(t)\hat{\mu}(\mathrm{d}t)\right) = \|\hat{\mu}\|_{\mathrm{TV}} \quad . \tag{15}$$

**Remark 2** (The constant dual polynomial issue)**.** *If the associated dual polynomial* $\hat{p}$ *is not constant, the support of* $\hat{\mu}$ *is finite, and is included in the set of its derivative roots, so the measure solution can be written as* $\hat{\mu} = \sum_{j=1}^{\hat{s}} \hat{a}_j \delta_{\hat{t}_j}$*. This follows from* (15)*.*

Equivalently, Equation (13) reads as follows

$$\forall t \in \mathbb{T}, \quad \frac{1}{n}\sum_{k=-f_c}^{f_c} (y_k - c_k(\hat{\mu}))\exp(2\pi\imath kt) = \hat{\lambda}\hat{p}(t) \quad . \tag{16}$$

## 3.2 No-overfitting and root-finding issues

In the sequel, we tackle the "*no-overfitting*" (see Section 2.1) and the "*constant dual polynomial*" issues. The "*constant dual polynomial*" issue is due to the use of root-finding algorithm which requires finding roots of the dual polynomial derivative. A practical limitation occurs when the dual polynomial is constant, in this case we cannot localize the primal solution support, which we refer to as the "*constant dual polynomial*" issue. We summarize our results from Propositions 6 and 7 in Figure 1. We may use the estimator (17) referred to as "Beurling Minimal Extrapolation" (BME for short) by [18], which extends the basis pursuit [16] in our context. The no-overfitting property is guaranteed by the following proposition.

**Proposition 6.** *Defining* $\lambda_{\min}(y) = 1/(\|\hat{c}^{(\mathrm{BME})}\|_2\sqrt{n})$ *and the problem*

$$\hat{\mu}^{(\mathrm{BME})} \in \underset{\mathcal{F}_n(\mu)=y}{\arg\min} \|\mu\|_{\mathrm{TV}} \ , \tag{17}$$

*and its dual formulation*

$$\begin{aligned} \hat{c}^{(\mathrm{BME})} \in \ & \underset{c\in\mathbb{C}^n}{\arg\max} \langle y, c\rangle \\ & s.t. \ \|\mathcal{F}_n^*(c)\|_\infty \leq 1 \ . \end{aligned} \tag{18}$$

*the following statements are equivalent*

*(i)* $\lambda \in ]0, \lambda_{\min}(y)]$*,*

*(ii)* $\hat{c} = \hat{c}^{(\mathrm{BME})}$*,*

*(iii)* $\hat{\sigma} = 0$ *(overfitting).*

*Remark that* $\lambda_{\min}(y) = 1/(\|\hat{c}^{(\mathrm{BME})}\|_2\sqrt{n}) > 1/\sqrt{n}$*.*

*Proof.* (i) ⇒ (ii): Choose $\lambda \in ]0, \lambda_{\min}(y)]$. Note that $n\lambda^2\|\hat{c}^{(\mathrm{BME})}\|_2 \leq n(\lambda_{\min}(y))^2\|\hat{c}^{(\mathrm{BME})}\|_2 \leq 1$. Hence, $\hat{c}^{(\mathrm{BME})} \in \mathcal{D}_n$, and since $\mathcal{D}_n \subset \{c \in \mathbb{C}^n : \|\mathcal{F}_n^*(c)\|_\infty \leq 1\}$, then $\hat{c} = \hat{c}^{(\mathrm{BME})}$.

(ii) ⇒ (iii): Assume that $\hat{c} = \hat{c}^{(\mathrm{BME})}$, then $y = n\hat{\lambda}\hat{c}^{(\mathrm{BME})} + \mathcal{F}_n(\hat{\mu})$ thanks to Eq. (11) and $\mathcal{F}_n(\hat{\mu}^{(\mathrm{BME})}) = y$ thanks to Eq. (17). Moreover, $\langle y, \hat{c}^{(\mathrm{BME})}\rangle = \|\hat{\mu}\|_{\mathrm{TV}}$ and $\lambda\langle y, \hat{c}\rangle = \frac{1}{2n\hat{\sigma}}\|y - \mathcal{F}_n(\hat{\mu})\|^2 + \frac{\hat{\sigma}}{2} + \lambda\|\hat{\mu}\|_{\mathrm{TV}}$ by strong duality. The only way the last equation holds is when $\hat{\sigma} = 0$ and that $y = \mathcal{F}_n(\hat{\mu})$.

(iii) ⇒ (i): Assume that $\hat{\sigma} = 0$, this leads to $\hat{\lambda} = 0$ thanks to the definition of $\hat{\lambda}$ below (11). Thanks to Eq. (11), $y = \mathcal{F}_n(\hat{\mu})$. This means that $(\hat{\mu}, \hat{\sigma})$ is solution of the problem

$$(\hat{\mu}, \hat{\sigma}) \in \underset{\substack{(\mu,\sigma)\in E^*\times\mathbb{R}_{++} \\ y=\mathcal{F}_n(\mu)}}{\arg\min} \frac{1}{2n\sigma}\|y - \mathcal{F}_n(\mu)\|_2^2 + \frac{\sigma}{2} + \lambda\|\mu\|_{\mathrm{TV}} \ .$$

and so

$$\hat{\mu} \in \underset{\substack{\mu\in E^* \\ y=\mathcal{F}_n(\mu)}}{\arg\min} \lambda\|\mu\|_{\mathrm{TV}} \ .$$

*i.e.,* $\hat{\mu} = \hat{\mu}^{(\mathrm{BME})}$.

By strong duality in Problem (4), one has $\lambda\|\hat{\mu}\|_{\mathrm{TV}} = \lambda\langle\hat{c}, y\rangle$ and by strong duality in Problem (18), $\lambda\|\hat{\mu}^{(\mathrm{BME})}\|_{\mathrm{TV}} = \lambda\langle\hat{c}^{(\mathrm{BME})}, y\rangle$. Hence $\langle\hat{c}, y\rangle = \langle\hat{c}^{(\mathrm{BME})}, y\rangle$ and one can choose $\hat{c}^{(\mathrm{BME})}$ as a dual optimal solution for Problem (10). So $\|\hat{c}^{(\mathrm{BME})}\|_2^2 \leq 1/(n\lambda^2)$, and (i) holds by definition of $\lambda_{\min}$.

We now proved the last statement of the proposition. Since $\|\hat{p}\|_\infty \leq 1$, Parseval's inequality leads to $\|\hat{c}\|_2 \leq 1$. If $\lambda < 1/\sqrt{n}$ then $\lambda^2 n\|\hat{c}\|_2 \leq \lambda^2 n < 1$, this means that the $\ell_2$ constraint in the dual formulation (10) is not saturated. With the first part of the proof, we deduce the result by choosing $\hat{c} = \hat{c}^{(\mathrm{BME})}$. Using (ii) ⇔ (i), one has $\lambda_{\min}(y) \geq 1/\sqrt{n}$. □

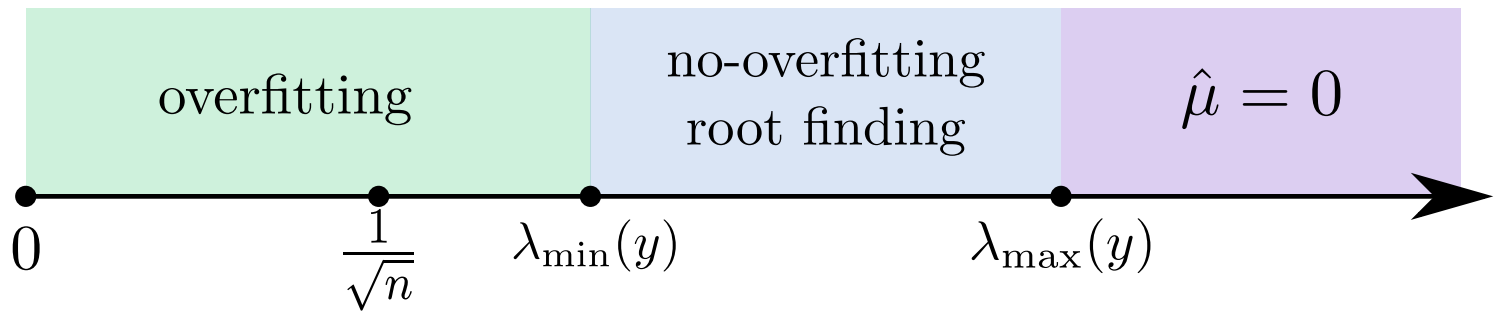

Figure 1 – CBLasso regimes for different value of the regularization parameter $\lambda$. When $\lambda \leq \lambda_{\min}(y)$, there is overfitting. When $\lambda_{\min}(y) < \lambda \leq \lambda_{\max}(y)$, the dual polynomial is not of constant modulus, so that the root finding can be done. We also show that the no-overfitting assumption holds in such a regime. Finally, when $\lambda > \lambda_{\max}(y)$, the solution is degenerated and $\hat{\mu} = 0$.

**Remark 3.** *As for the BLasso defined by solving*

$$\hat{\mu}^{\text{BLasso}} \in \underset{\mu \in E^*}{\arg\min} \frac{1}{2n} \|y - \mathcal{F}_n(\mu)\|_2^2 + \lambda \|\mu\|_{\text{TV}}, \tag{19}$$

*note that if $\lambda$ is chosen large enough, 0 is the unique solution of the CBLasso problem given by* (4)*, the threshold being $\lambda_{\max}(y) = \|\mathcal{F}_n^*(y)\|_\infty / (\sqrt{n}\|y\|_2)$ (for the BLasso it is simply $\|\mathcal{F}_n^*(y)\|_\infty / n$). This result is easily deduced thanks to the KKT conditions.*

The next proposition ensures that the root-finding is always possible, meaning that when the primal solution is non-zero, then the dual polynomial is non-constant.

**Proposition 7.** *For $\lambda \in ]\lambda_{\min}(y); \lambda_{\max}(y)]$, the polynomial $|\hat{p}|^2$ is non-constant.*

*Proof.* See Appendix E.4. □

### 3.3 Semi Definite Program formulation of the CBLasso

We write $A \succcurlyeq 0$ when a symmetric matrix $A$ is semi-definite positive. Let us recall a classical property expressing the CBLasso as a semi-definite program (SDP), see [21, Sec. 4.3] or [14, 37] for instance.

**Proposition 8.** *For any $c \in \mathbb{C}^n$, the following holds*

$$\|\mathcal{F}_n^* c\|_\infty^2 \leq 1 \Leftrightarrow \exists \Lambda \in \mathbb{C}^{n \times n} \text{ satisfying } \Lambda^* = \Lambda \text{ and } \begin{cases} \begin{pmatrix} \Lambda & c \\ c^* & 1 \end{pmatrix} \succcurlyeq 0 \ , \\ \sum_{i=1}^{n-j+1} \Lambda_{i,i+j-1} = \delta_{j,1}, \forall j \in [n] \ . \end{cases} \tag{20}$$

*where $\delta_{k,l}$ is the standard Kronecker symbol.*

Remark that $A \succcurlyeq 0$ and $B \succcurlyeq 0$ is equivalent to $\begin{pmatrix} A & 0 \\ 0 & B \end{pmatrix} \succcurlyeq 0$. From properties of the Schur complement (*cf.* [11, p. 651]) a block matrix $\begin{pmatrix} A & B \\ B^* & C \end{pmatrix} \succcurlyeq 0 \Leftrightarrow A \succcurlyeq 0$ and $C - B^* A^{-1} B \succcurlyeq 0$.

Applying this, one can represent the dual feasible set $\mathcal{D}_n$, as an SDP condition and the dual problem can be cast as follows

$$\max_{c\in\mathbb{C}^n} \lambda \langle y, c\rangle \quad \text{such that} \quad \begin{cases} \begin{pmatrix} \Lambda & c \\ c^* & 1 \end{pmatrix} \succcurlyeq 0 \ , \\ \sum_{i=1}^{n-j+1} \Lambda_{i,i+j-1} = \delta_{j,1}, \forall j \in [n] \ , \\ \begin{pmatrix} \mathrm{Id}_n & \lambda\sqrt{n}c \\ \lambda\sqrt{n}c^* & 1 \end{pmatrix} \succcurlyeq 0. \end{cases} \tag{21}$$

## 3.4 From the dual to the primal

By solving Problem (21), one can identify $\hat{p}$, the dual polynomial (12), and the set of locations where the latter reaches unit modulus and in which the support of $\hat{\mu}$ is included, see Remark 2. We recall that the support of all the solutions of Problem (4) are included in the level set $|\hat{p}|^2 = 1$. Once this set is identified, remark that solutions to Problem (4) are equivalently solutions to a finite dimensional one:

$$(\hat{\mu}, \hat{\sigma}) \in \underset{(\mu,\sigma)\in\hat{E}^*\times\mathbb{R}_{++}}{\arg\min} \frac{1}{2n\sigma}\|y - \mathcal{F}_n(\mu)\|_2^2 + \frac{\sigma}{2} + \lambda\|\mu\|_{\mathrm{TV}} \ , \tag{22}$$

where $\hat{E}^* := (\mathcal{C}(\mathrm{supp}(\hat{\mu}), \mathbb{C}), \|\cdot\|_\infty)$ the space of Borelian measure, whose support is included in $\mathrm{supp}(\hat{\mu}) = \{\hat{t}_j, j = 1, \ldots, \hat{s}\}$, a set found thanks to the dual formulation from the previous section. Indeed, any solution $\hat{\mu}$ to Problem (4) belongs to $\hat{E}^*$ so that it is equivalently a solution to (22).

We can now introduce the design matrix $X \in \mathbb{R}^{n\times\hat{s}}$, defined by $X_{k,j} = \overline{\varphi_k}(\hat{t}_j)$. Considering the estimators $(\hat{a}, \hat{\sigma})$ defined by

$$(\hat{a}, \hat{\sigma}) \in \underset{(a,\sigma)\in\mathbb{R}^{\hat{s}}\times\mathbb{R}_{++}}{\arg\min} \frac{1}{2n\sigma}\|y - Xa\|_2^2 + \frac{\sigma}{2} + \lambda\|a\|_1 \ , \tag{23}$$

one can check that $(\hat{\mu}, \hat{\sigma})$ satisfies the original optimality condition for Problem (4), where

$$\hat{\mu} = \sum_{j=1}^{\hat{s}} \hat{a}_j \delta_{\hat{t}_j} \ .$$

To solve (23), we proceed following the alternate minimization procedure proposed in [36], that consists in alternating between a Lasso step and a noise level estimation step (*i.e.,* computing the norm of the residuals). Note that the Lasso step is simple in this case, since the KKT condition reads $X^*(Xa - y) + \lambda\hat{\sigma}\hat{\zeta} = 0$ where $\hat{\zeta} = \mathrm{sign}(X^*\hat{c})$. Provided that the matrix $X^*X$ can be stored and inverted, one can use

$$\hat{a} = X^+ y - \lambda\hat{\sigma}(X^*X)^{-1}\hat{\zeta} \tag{24}$$

along the iterative process.

### 3.5 Experiments

The source code of all the experiments presented in this section can be downloaded from this gitHub repository[2]; a notebook can also be found at this address[3]. First, let us summarize the description of the proposed algorithm: given the data $y \in \mathbb{C}^n$

1. Set $\lambda = \alpha\,\lambda_{\max}(y)$, for a constant $\alpha \in (0,1)$ fraction of $\lambda_{\max}(y) = \|\mathcal{F}^*(y)\|_\infty/(\sqrt{n}\|y\|_2)$;
2. Solve Problem (21) to find the coefficients $\hat{c}$ of the dual polynomial $\hat{p}$. For this step, we use the cvx Matlab toolbox [27, 28];
3. Identify $\mathrm{supp}(\hat{\mu})$ using the roots of $1 - |\hat{p}|^2$ and construct the matrix X described above;
4. Solve Problem (23) as follows: for an initial value of $\hat{\sigma}$, until some stopping criterion,
   (a) Solve Problem (23) using (24) with $\hat{\sigma}$ to compute $\hat{a}$,
   (b) Update $\hat{\sigma} = \|y - X\hat{a}\|_2/\sqrt{n}$ using the new value of $\hat{a}$,

In our experiments, we have chosen in Step 4 the stopping criterion combining (i) a maximal number of iterations fixed to 1000 and (ii) a tolerance threshold of $10^{-4}$ between two iterates of $\hat{\sigma}$.

**Measure estimation** We run this algorithm for estimating a 3-spikes measure. The measure is generated by drawing uniformly at random support locations in the torus satisfying the separation condition. The spike amplitudes are set to 1 or $-1$ at random. The noise level $\sigma_0$ is fixed to 1 and $n = 161$. For this experiment we fix $\lambda$ to be equal to $\lambda_{\max}(y)/2 = \|\mathcal{F}_n^*(y)\|_\infty/(2\sqrt{n}\|y\|_2)$. The results are presented in Fig. 2 and compared with a BLasso approach (solved thanks to (24) for $\lambda = n\lambda_{\max}^{\text{BLasso}}(y)/2$ with $\lambda_{\max}^{\text{BLasso}}(y) = \|\mathcal{F}_n^*(y)\|_\infty/n$ and $\hat{\sigma} = \sigma_0$, the true level of noise) over the estimated support $\mathrm{supp}(\hat{\mu})$. First, note that both the BLasso and the CBLasso methods can recover the true support $\mathrm{supp}(\mu^0)$. Second, the CBLasso better estimates the spikes magnitude in the original measure than the BLasso due to a better scaling of the regularizing factor.

**Noise estimation** In order to illustrate noise estimation performance provided by the CBLasso method, we run the following experiment. Following the same procedure described above, we draw at random 100 target measures replica composed of 3 spikes with support satisfying the separation condition. For each target measure $\mu^0$, we observe $y = \mathcal{F}_n(\mu^0) + \varepsilon$ with $n = 161$ and $\varepsilon$ a complex Gaussian vector such that $\varepsilon \overset{(d)}{=} \varepsilon^{(1)} + i\varepsilon^{(2)}$ and $\varepsilon^{(1)}, \varepsilon^{(2)} \sim \mathcal{N}(0, (1/2)\,\mathrm{Id}_n)$ (here we choose $\sigma_0 = 1/\sqrt{2}$) and we perform the algorithm proposed above. In Fig. 3, we present a boxplot on the value $\hat{\sigma}$ for the 100 CBLasso estimations. One can remark that $\hat{\sigma}$ presents a bias compared to the noise level equal to 1, but this bias will decrease as $n$ increases. Indeed, Proposition 4 shows that $\hat{\sigma}$ is close to $\|\varepsilon\|_2/\sqrt{n}$ whose expectation is $\sqrt{2}\sigma_0\mathbb{E}\|g\|_2/\sqrt{2n}$ with $g$ standard Gaussian in dimension $2n$. We deduce that

$$\hat{\sigma} \simeq \sqrt{2}\sigma_0 \times \frac{\mathbb{E}\|g\|_2}{\sqrt{2n}} = \sqrt{2}\sigma_0 \times \frac{\Gamma(n+1/2)}{\sqrt{n}\Gamma(n)} \to \sqrt{2}\sigma_0\,,$$

showing that $\hat{\sigma}/\sqrt{2}$ is consistent estimator of $\sigma_0$.

[2] https://github.com/claireBoyer/CBLasso
[3] http://www.lsta.upmc.fr/boyer/codes/html_CBlasso_vs_Blasso/script_example1_CBlasso_vs_Blasso.html

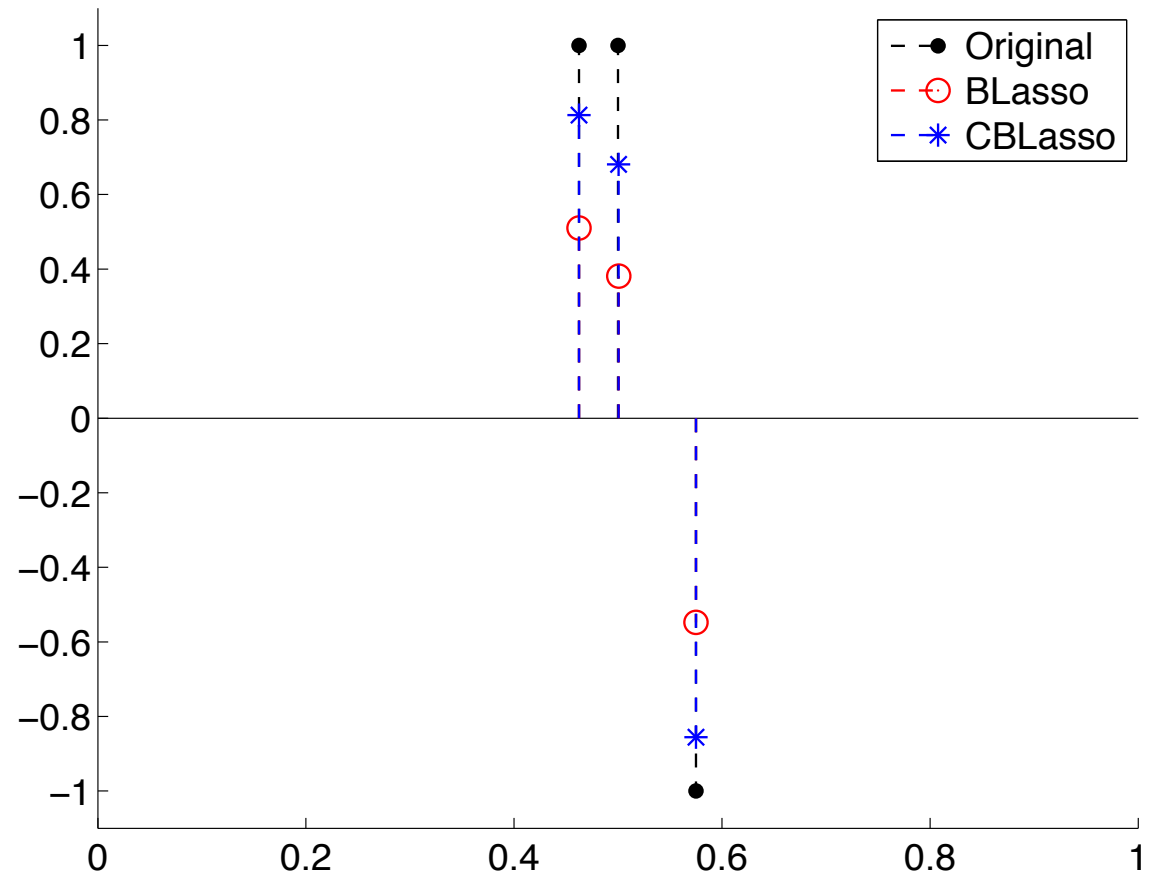

Figure 2 – Reconstruction of a discrete measure. The original measure $\mu^0$ is composed of 3 spikes (in black). The reconstructed measure $\hat{\mu}$ using our proposed CBLasso (in blue). In comparison, we plot the reconstructed measure using the BLasso, (in red).

In Fig. 3, we also compare the CBLasso noise estimation to

$$\hat{\sigma}^{\mathrm{BLasso}} = \frac{1}{\sqrt{n - \hat{s}^{\mathrm{BLasso}}}} \|y - \mathcal{F}_n(\hat{\mu}^{\mathrm{BLasso}})\|_2,$$

proposed in [34], in which $\hat{\mu}^{\mathrm{BLasso}}$ denotes the reconstructed measure that is supported on $\hat{s}^{\mathrm{BLasso}}$ spikes using the BLasso. The CBLasso approach provides a satisfactory noise level estimation w.r.t. the heuristics defined above.

# Acknowledgments

The authors are grateful to Sara van de Geer, particularly for introducing them into the techniques she presented during the ≪ École d'été de Probabilités de Saint-Flour XLV ≫.

# Références

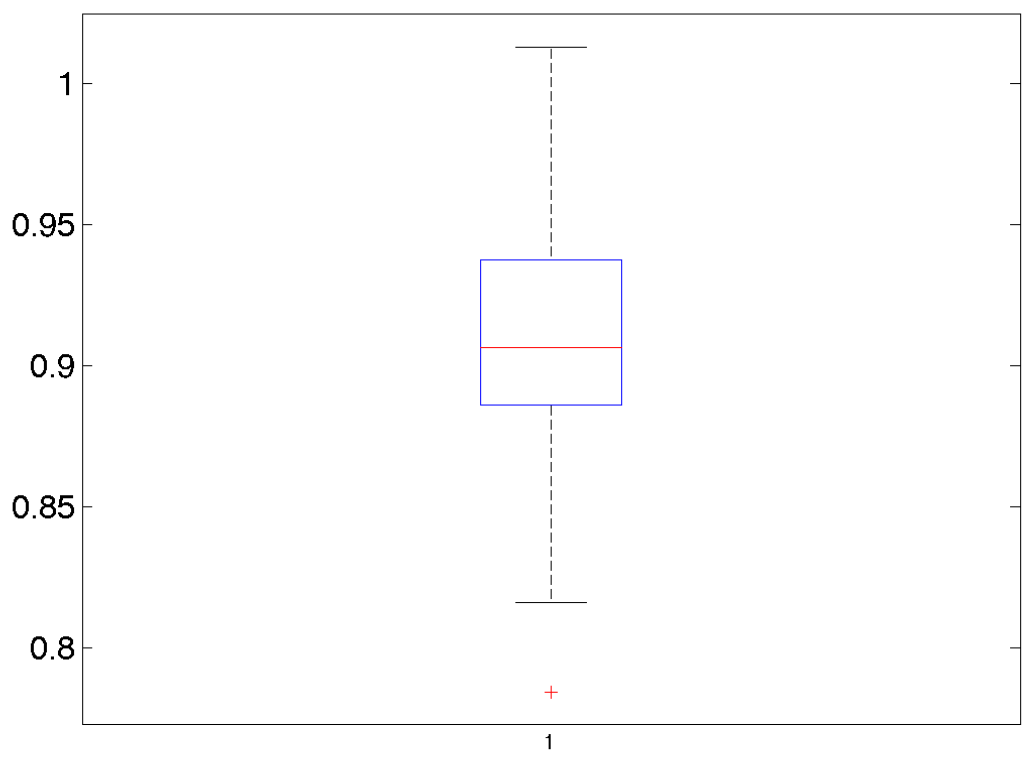

Figure 3 – Boxplot on $\hat{\sigma}$ for 100 CBLasso consistent estimations of $\sqrt{2}\sigma_0 = 1$. We compare our method to $\hat{\sigma}^{\text{BLasso}} = \|y - \mathcal{F}_n(\hat{\mu}^{\text{BLasso}})\|_2/\sqrt{n - \hat{s}^{\text{BLasso}}}$ proposed in [34] where $\hat{\mu}^{\text{BLasso}}$ is the reconstructed measure supported on $\hat{s}^{\text{BLasso}}$ spikes via BLasso. Noise estimation using CBLasso is clearly closer to $\sigma_0$ than $\hat{\sigma}^{\text{BLasso}}$.

# Reading guide

The proofs of this paper are quite long and may be difficult to follow. The main part is devoted for proving Theorem 1 as we need the control of the prediction error to derive the result on estimation of the target measure, namely Theorem 2. More precisely, we need (33) that controls the Bregman divergence of the TV norm at point $(\hat{\mu}, \mu^0)$. If we admit this control, the proof of Theorem 2 is only two pages and partially follows from [2, 25]. Regarding this proof, the main contribution is to control the influence of the noise estimation error on $\hat{\sigma}$, bounding it from above and/or below in each step. This is new since this paper is the first to address the noise estimation issue in such a context. This proof is presented in Section B. Three ingredients are part of Theorem 1 and Theorem 2 proofs.

1. The first one is the noise level control, namely bounding the probability of $\{\|\mathcal{F}_n^*(\varepsilon)\|_\infty \leq \lambda\}$ and $\{\|\mathcal{F}_n^*(\varepsilon)\|_\infty/\sqrt{n}\|\varepsilon\|_2 \leq R\}$. This step required new results on the supremum of processes indexed by the torus. This is done using the Rice formula [3, Proposition 4.1, Page 93] presented in Section D.2

2. One recent breakthrough in Super-Resolution has been brought by pioneering constructions [14, 18, 9] of *dual polynomials*, namely proving the existence of convenient sub-gradients of the TV norm at the target measure (and, hence, assuming the separation condition of Assumption 2). These constructions are now well referenced and we omit their proofs in this paper. They are briefly synthesized though in Section C.

3. The last ingredient is a mixture of optimality conditions derived from convexity (see Section E) and simple but non trivial *ad hoc* inequalities.

The proofs presentation is essentially focused on these last points since we believe that they are specific to and at the heart of the Super Resolution framework. As we have seen, these *ad hoc* inequalities steps are presented in Section B for Theorem 2.

As for Theorem 1, this last ingredient is costly, requiring several pages, see Section A. Its proof is based on the pioneering paper [37]. However, the proof presented here, on the prediction error, differs from their since we take into account the noise estimation. It changes large parts of the proof of [37] and Section A is devoted to this task. In a nutshell, the noise estimation is given by $\widehat{\lambda} = \lambda\hat{\sigma}$ where $\lambda$ is a tuning parameter of our algorithm, while the noise level $\|\mathcal{F}_n^*(\varepsilon)\|_\infty$ can be bounded (with large probability) by $\tilde{\lambda}$ in the proof, see for instance Lemma 10. Observe that $\lambda$ itself has been tuned so that it bounds the noise level but we need to draw a (technical) distinction here. Doing so, we are able to assess the probability of some key events such that $\{\widetilde{\lambda}/\widehat{\lambda} \leq (C_F \wedge C_N)/(2C)\}$, see Page 22. This event has been controlled in [37] with "high probability" while "choosing large enough constants" which was rightfully enough for the purpose of [37]. In this paper, we carefully quantify these assertions as done in Lemma 12. The second main difference with the proof of [37] relies on the fact that we have to track the noise estimation error in all the *ad hoc* inequalities steps. This has been achieved tuning $\lambda$, $\tilde{\lambda}$, $\widehat{\lambda}$, $\tilde{\beta}$, $\widehat{\beta}$, etc., in a suitable manner. The proofs have been organized through the paper according to the aforementioned remarks.

# A Proof of Theorem 1

Let us define $c_k(\hat{\mu} - \mu^0) = \int_{\mathbb{T}} \exp(-2\pi\imath kt)(\hat{\mu} - \mu^0)(\mathrm{d}t)$ for $-f_c \leq k \leq f_c$, and introduce the trigonometric polynomial $\Delta$ of degree $f_c$ defined by

$$\Delta = \sum_{|k| \leq f_c} c_k(\hat{\mu} - \mu^0)\varphi_k \ , \tag{25}$$

with $\varphi_k(\cdot) = \exp(2\pi\imath k \cdot)$. One can write

$$\frac{1}{n}\|\mathcal{F}_n(\hat{\mu} - \mu^0)\|_2^2 = \frac{1}{n}\Re\Big[\int_{\mathbb{T}} \sum_{|k|\leq f_c} \overline{c_k(\hat{\mu} - \mu^0)} e^{-2\imath\pi kt}(\hat{\mu} - \mu^0)(\mathrm{d}t)\Big] = \frac{1}{n}\Re\Big[\int_{\mathbb{T}} \bar{\Delta}(t)(\hat{\mu} - \mu^0)(\mathrm{d}t)\Big] \ .$$

Applying Lemma 9 (see Section A.1 below) with $q = \bar{\Delta}$ and $\nu = \hat{\mu} - \mu^0$, one gets

$$\begin{aligned}
\Re\Big[\int_{\mathbb{T}} \bar{\Delta}(\hat{\mu} - \mu^0)(\mathrm{d}t)\Big] &\leq \|\Delta\|_\infty \left[ \sum_{i=1}^{s_0} \Bigg( \underbrace{\left|\int_{N_i} \nu(\mathrm{d}t)\right|}_{I_0^i} + \underbrace{n\left|\int_{N_i} (t - t_i)\nu(\mathrm{d}t)\right|}_{I_1^i} + \underbrace{\frac{n^2}{2}\left|\int_{N_i} (t - t_i)^2\nu(\mathrm{d}t)\right|}_{I_2^i} \Bigg) \right. \\
&\qquad \left. + \int_F |\nu|(\mathrm{d}t) \right] \ , \\
&\leq \|\Delta\|_\infty \left[ \int_F |\nu|(\mathrm{d}t) + I_0 + I_1 + I_2 \right] \ ,
\end{aligned}$$

with $I_j = \sum_{i=1}^{s_0} I_j^i$ for $j = 0, 1, 2$. Therefore,

$$\frac{1}{n}\|\mathcal{F}_n(\hat{\mu} - \mu^0)\|_2^2 \leq \frac{1}{n}\|\Delta\|_\infty \left[ \int_F |\nu|(\mathrm{d}t) + I_0 + I_1 + I_2 \right] \ . \tag{26}$$

The result in Theorem 1 follows by bounding each term in (26) using Lemmas 10, 11 and 12, presented in the sequel.

## A.1 Preliminary lemma

**Lemma 9.** *For all trigonometric polynomial $q$ of degree less than $f_c$ and for all $(t_i^0)_{1\leq i\leq s_0} \in \mathbb{T}^{s_0}$ satisfying the separation condition given in Assumption 2, we have for any $\nu \in E^*$,*

$$\left|\int_{\mathbb{T}} q\mathrm{d}\nu\right| \leq \|q\|_\infty \left[ \int_F |\nu|(\mathrm{d}t) + \sum_{i=1}^{s_0} \left( \left|\int_{N_i} \nu(\mathrm{d}t)\right| + n\left|\int_{N_i} (t - t_i^0)\nu(\mathrm{d}t)\right| + \frac{n^2}{2}\left|\int_{N_i} (t - t_i^0)^2\nu(\mathrm{d}t)\right| \right) \right] \ ,$$

*where $N_j$ and $F$ are defined in (6) and (7).*

*Proof.* Given the definitions of $F$ and the $(N_i)$'s, one can write

$$\left|\int_{\mathbb{T}} q(t)\nu(\mathrm{d}t)\right| \leq \left|\int_F q(t)\nu(\mathrm{d}t)\right| + \sum_{i=1}^{s_0} \left|\int_{N_i} q(t)\nu(\mathrm{d}t)\right| \leq \|q\|_\infty \int_F |\nu|(\mathrm{d}t) + \sum_{i=1}^{s_0} \left|\int_{N_i} q(t)\nu(\mathrm{d}t)\right| \ .$$

In the sequel, we may identify $\mathbb{T}$ with $\mathbb{R}/\mathbb{Z}$ using $[-1/2, 1/2)$ as fundamental polygon. Using the Taylor-Lagrange expansion

$$\left|q(t) - q(t_i^0) - q'(t_i^0)(t - t_i^0)\right| \leq \|q''\|_\infty \frac{(t - t_i^0)^2}{2} \ ,$$

and the Bernstein inequality [10, Theorem 5.1.4] for trigonometric polynomials (reminding $f_c \leq n$)

$$\|q'\|_\infty \leq n\|q\|_\infty \quad \text{and} \quad \|q''\|_\infty \leq n^2\|q\|_\infty \ ,$$

we can derive for a fixed $i \in \{1, \ldots, s_0\}$

$$\begin{aligned}\left|\int_{N_i} q(t)(\nu)(\mathrm{d}t)\right| \leq \left|q(t_i^0)\right| \left|\int_{N_i} (\hat{\mu} - \mu^0)(\mathrm{d}t)\right| + |q'(t_i^0)| \left|\int_{N_i} (t - t_i^0)(\hat{\mu} - \mu^0)(\mathrm{d}t)\right| \\ + \frac{n^2\|q\|_\infty}{2} \int_{N_i} (t - t_i^0)^2 |\nu|(\mathrm{d}t) \ , \\ \leq \|q\|_\infty \left(\left|\int_{N_i} (\nu)(\mathrm{d}t)\right| + n\left|\int_{N_i} (t - t_i^0)(\nu)(\mathrm{d}t)\right| + \frac{n^2}{2}\left|\int_{N_i} (t - t_i^0)^2(\nu)(\mathrm{d}t)\right|\right) \ ,\end{aligned}$$

as claimed. □

## A.2 Control of $\|\Delta\|_\infty$

**Lemma 10.** *Let $\widetilde{\alpha} \in (0, 1)$ and set $\widetilde{\lambda} := 2\sigma_0\sqrt{\log(n/\widetilde{\alpha})/n}$. Then, reminding $\widehat{\lambda} = \lambda\hat{\sigma}$, it holds*

$$\|\Delta\|_\infty = \sup_{t\in\mathbb{T}} \left|\sum_{|k|\leq f_c} c_k(\hat{\mu} - \mu^0)\varphi_k(t)\right| \leq n\left(\widetilde{\lambda} + \widehat{\lambda}\right) \ ,$$

*with probability greater than $1 - \widetilde{\alpha}$.*

*Proof.* Recall that $\mathcal{F}_n(\mu^0) = y - \varepsilon$ to get

$$\begin{aligned}\|\Delta\|_\infty &\leq \sup_{t\in\mathbb{T}} \left|\sum_{|k|\leq f_c} \varepsilon_k\varphi_k(t)\right| + \sup_{t\in\mathbb{T}} \left|\sum_{|k|\leq f_c} (y_k - c_k(\hat{\mu}))\varphi_k(t)\right| \ , \\ &\leq \|\mathcal{F}_n^*(\varepsilon)\|_\infty + \sup_{t\in\mathbb{T}} \left|\sum_{|k|\leq f_c} (y_k - c_k(\hat{\mu}))\varphi_k(t)\right| \ .\end{aligned}$$

Using Lemma 18, it holds that $\|\mathcal{F}_n^*(\varepsilon)\|_\infty \leq n\widetilde{\lambda}$ with probability greater than $1 - \widetilde{\alpha}$. Using the KKT conditions (13), we have that

$$\sup_{t\in\mathbb{T}} \left|\sum_{|k|\leq f_c} (y_k - c_k(\hat{\mu}))\varphi_k(t)\right| \leq \widehat{\lambda}\|n\hat{p}\|_\infty \leq n\widehat{\lambda} \ .$$

We deduce the result from this last point. □

## A.3 Control of $I_0$ and $I_1$ by $I_2 + \int_F |\nu|(\mathrm{d}t)$

**Lemma 11.** *There exists a numerical constant $C_2 > 0$ such that*

$$I_0 \le C_2 \left( s_0 \widehat{\lambda} + I_2 + \int_F |\nu|(\mathrm{d}t) \right) \quad \text{and} \quad I_1 \le C_2 \left( s_0 \widehat{\lambda} + I_2 + \int_F |\nu|(\mathrm{d}t) \right) \ .$$

*Proof.* Invoke Lemma 2 of [38] to get the result. □

## A.4 Control of $I_2 + \int_F |\nu|(\mathrm{d}t)$

Setting $\widehat{\alpha} = n^{-\widehat{\beta}}$ and $\widetilde{\alpha} = n^{-\widetilde{\beta}}$ for some well chosen constants $\widehat{\beta} > 0, \widetilde{\beta} > 0$ (to be specified later), the proof reduces to control the quantity

$$\int_F |\hat{\mu} - \mu^0|(\mathrm{d}t) + I_2 = \int_F |\hat{\mu} - \mu^0|(\mathrm{d}t) + \sum_{i=1}^{s_0} I_2^i = \int_F |\hat{\mu} - \mu^0|(\mathrm{d}t) + \frac{n^2}{2} \sum_{i=1}^{s_0} \left| \int_{N_i} (t - t_i)^2 (\hat{\mu} - \mu^0)(\mathrm{d}t) \right| \ ,$$

and $\widehat{\lambda}$ in Lemma 10.

**Lemma 12.** *Under the assumptions of Theorem 1, with a probability of failure that decays as a power of $n$, it holds*

$$\int_F |\nu|(\mathrm{d}t) + I_2 \le \frac{3\sqrt{2}}{2} \left( 1 + \frac{C_1}{C} + \frac{2C_1}{C_F \wedge C_N} \right) s_0 \sigma_0 \lambda \ .$$

*where the constants $C_1, C_N, C_F > 0$ are defined in Section C and $C > 0$ is an universal constant.*

*Proof.* The proof follows several steps.
**Preliminary** First, let us fix $S := \mathrm{supp}(\mu^0) = \{t_1^0, \dots, t_{s_0}^0\}$ and then let $\Pi_S : E^* \to E^*$ be defined for $\nu \in E^*$, such that $\Pi_S(\nu)$ is the atomic part of $\nu$ on $S$. Considering that $\nu = \hat{\mu} - \mu^0$ and using the triangle inequality, one can write that

$$\|\Pi_S(\nu)\|_{\mathrm{TV}} \le \left| \int_{\mathbb{T}} q_1(t) \nu(\mathrm{d}t) \right| + \left| \int_{\mathbb{T}} q_1(t) \nu_{|S^c}(\mathrm{d}t) \right| \ ,$$

where $\nu_{|S^c} = \nu - \Pi_S(\nu)$, and $q_1$ is the interpolation polynomial interpolating the phases of $\Pi_S(\nu)$, defined as in Lemma 13. By Hölder's inequality and Lemma 10, one has

$$\left| \int_{\mathbb{T}} q_1(t) \nu(\mathrm{d}t) \right| \le \|q_1\|_1 \|\Delta\|_\infty \le \|q_1\|_1 (\widetilde{\lambda} + \widehat{\lambda}) n \ .$$

By noticing the following disjoint union $\mathbb{T} = F \sqcup \left( \cup_{i=1}^{s_0} N_i \setminus \{t_i^0\} \right) \sqcup S$, we deduce that

$$\|\Pi_S(\nu)\|_{\mathrm{TV}} \le \|q_1\|_1 (\widetilde{\lambda} + \widehat{\lambda}) n + \left| \int_F q_1(t) \nu_{|S^c}(\mathrm{d}t) \right| + \sum_{i=1}^{s_0} \left| \int_{N_i \setminus \{t_i^0\}} q_1(t) \nu_{|S^c}(\mathrm{d}t) \right| + \left| \int_S q_1(t) \nu_{|S^c}(\mathrm{d}t) \right| \ ,$$

in which the last term is equal to 0 since $\nu_{|S^c}$ has no mass on $S$. Note that, for any borelian $A \subseteq \mathbb{T}$, it holds $\nu_{|S^c}(F \cap A) = \nu(F \cap A)$. Therefore,

$$\|\Pi_S(\nu)\|_{\mathrm{TV}} \le \|q_1\|_1 (\widetilde{\lambda} + \widehat{\lambda}) n + \left| \int_F q_1(t) \nu(\mathrm{d}t) \right| + \sum_{i=1}^{s_0} \left| \int_{N_i \setminus \{t_i\}} q_1(t) \nu_{|S^c}(\mathrm{d}t) \right| \ .$$

By Lemma 13, one can write that

$$\|\Pi_S(\nu)\|_{\mathrm{TV}} \le \|q_1\|_1(\widetilde{\lambda}+\widehat{\lambda})n + (1-C_F)\int_F |\nu|(\mathrm{d}t) + \sum_{i=1}^{s_0}\left|\int_{N_i\setminus\{t_i^0\}} q_1(t)\nu_{|S^c}(\mathrm{d}t)\right| \ .$$

Moreover, using Lemma 13(ii),

$$\begin{aligned}\left|\int_{N_i\setminus\{t_i^0\}} q_1(t)\nu_{|S^c}(\mathrm{d}t)\right| &\le \int_{N_i\setminus\{t_i^0\}} |q_1|(t)|\nu|_{|S^c}(\mathrm{d}t) \le \int_{N_i\setminus\{t_i^0\}} \left(1-\frac{C_N}{2}n^2(t-t_i^0)^2\right)|\nu|(\mathrm{d}t) \ ,\\ &\le \int_{N_i\setminus\{t_i^0\}} |\nu|(\mathrm{d}t) - C_N I_2^i \ .\end{aligned}$$

Combining the last two inequalities, one gets

$$\begin{aligned}\|\Pi_S(\nu)\|_{\mathrm{TV}} &\le \|q_1\|_1(\widetilde{\lambda}+\widehat{\lambda})n + \|\nu_{|S^c}\|_{\mathrm{TV}} - C_F\int_F |\nu|(\mathrm{d}t) - C_N\sum_{i=1}^{s_0} I_2^i \ ,\\ &\le \|q_1\|_1(\widetilde{\lambda}+\widehat{\lambda})n + \|\nu_{|S^c}\|_{\mathrm{TV}} - C_F\int_F |\nu|(\mathrm{d}t) - C_N I_2 \ ,\end{aligned}$$

and finally,

$$\|\Pi_S(\nu)\|_{\mathrm{TV}} - \|\nu_{|S^c}\|_{\mathrm{TV}} \le -C_F\int_F |\nu|(\mathrm{d}t) - C_N I_2 + \|q_1\|_1(\widetilde{\lambda}+\widehat{\lambda})n \ . \tag{27}$$

**Trade-off between $\widehat{\lambda}$ and $\widetilde{\lambda}$.** Secondly, by optimality of $\hat{\mu}$, we have

$$\frac{1}{2n}\|y-\mathcal{F}_n(\hat{\mu})\|_2^2 + \widehat{\lambda}\|\hat{\mu}\|_{\mathrm{TV}} \le \frac{1}{2n}\|\varepsilon\|_2^2 + \widehat{\lambda}\|\mu^0\|_{\mathrm{TV}} \ ,$$

then, invoking Lemmas 9 and 11 (at the last step), one has

$$\begin{aligned}n\widehat{\lambda}\left(\|\hat{\mu}\|_{\mathrm{TV}} - \|\mu^0\|_{\mathrm{TV}}\right) &\le \frac{1}{2}\left(\|\varepsilon\|_2^2 - \|y-\mathcal{F}_n(\hat{\mu})\|_2^2\right) \ ,\\ &= \frac{1}{2}\left(\|\varepsilon\|_2^2 - \|y-\mathcal{F}_n(\mu^0-\hat{\mu})+\varepsilon\|_2^2\right),\\ &= \frac{1}{2}\left(2\left\langle\varepsilon,\mathcal{F}_n(\hat{\mu}-\mu^0)\right\rangle - \|\mathcal{F}_n(\mu^0-\hat{\mu})\|_2^2\right) \ ,\\ &\le \left|\left\langle\varepsilon,\mathcal{F}_n(\hat{\mu}-\mu^0)\right\rangle\right| = \left|\Re\left(\int_{\mathbb{T}}\overline{\mathcal{F}_n^*(\varepsilon)}\mathrm{d}\nu\right)\right| \ ,\\ &\le n\widetilde{\lambda}\cdot C\left(s_0\widehat{\lambda} + I_2 + \int_F |\nu|(\mathrm{d}t)\right) \ ,\end{aligned} \tag{28}$$

for some universal constant $C>0$. Considering the triangle inequality coupled with the separability property of $\|\cdot\|_{\mathrm{TV}}$, one has $\|\hat{\mu}\|_{\mathrm{TV}} = \|\mu^0+\nu\|_{\mathrm{TV}} \ge \|\mu^0\|_{\mathrm{TV}} - \|\Pi_S(\nu)\|_{\mathrm{TV}} + \|\nu_{|S^c}\|_{\mathrm{TV}}$. It yields

$$n\widehat{\lambda}\left(\|\nu_{|S^c}\|_{\mathrm{TV}} - \|\Pi_S(\nu)\|_{\mathrm{TV}}\right) \le n\widehat{\lambda}\left(\|\hat{\mu}\|_{\mathrm{TV}} - \|\mu^0\|_{\mathrm{TV}}\right) \le n\widetilde{\lambda}C\left(s_0\widehat{\lambda} + I_2 + \int_F |\nu|(\mathrm{d}t)\right) \ . \tag{29}$$

Combining (27) and (29), we finally get

$$C_F \int_F |\nu|(\mathrm{d}t) + C_N I_2 - \|q_1\|_1(\widetilde{\lambda} + \widehat{\lambda})n \leq \frac{\widetilde{\lambda}}{\widehat{\lambda}} C \left( s_0 \widehat{\lambda} + I_2 + \int_F |\nu|(\mathrm{d}t) \right) \ .$$

Using Lemma 15, one has

$$(C_F \wedge C_N) \left( \int_F |\nu|(\mathrm{d}t) + I_2 \right) \leq \frac{\widetilde{\lambda}}{\widehat{\lambda}} C \left( s_0 \widehat{\lambda} + I_2 + \int_F |\nu|(\mathrm{d}t) \right) + C_1 s_0 (\widetilde{\lambda} + \widehat{\lambda}) \ ,$$

On the event $\left\{ \widetilde{\lambda}/\widehat{\lambda} \leq (C_F \wedge C_N)/(2C) \right\}$, then

$$\int_F |\nu|(\mathrm{d}t) + I_2 \leq \left( 1 + \frac{C_1}{C} + \frac{2C_1}{C_F \wedge C_N} \right) s_0 \widehat{\lambda} \ . \tag{30}$$

**Control of the event $\left\{ \widetilde{\lambda}/\widehat{\lambda} \leq (C_F \wedge C_N)/(2C) \right\}$.** Recall that $\widetilde{\lambda} := 2\sigma_0 \sqrt{\log(n/\widetilde{\alpha})/n}$ has been chosen so that $\mathbb{P}\left\{ \|\mathcal{F}_n^*(\varepsilon)\|_\infty \leq n\widetilde{\lambda} \right\} \geq 1 - \widetilde{\alpha}$ , using Lemma 18. Moreover, on the events $\left\{ \frac{\|\mathcal{F}_n^*(\varepsilon)\|_\infty}{\sqrt{n}\|\varepsilon\|_2} \leq R \right\}$ and $\left\{ \frac{\|\varepsilon\|_2}{\sqrt{n}} \geq \underline{\sigma} \right\}$ with the choice $R = \sqrt{2\log(n/\widehat{\alpha})/n}$ and $\underline{\sigma} = \sqrt{2}\sigma_0 \left( 1 - \sqrt{-2\log\widehat{\alpha}/n} \right)^{1/2}$, one can invoke Proposition 4 with $\eta = 1/2$ to obtain

$$\widehat{\lambda} = \hat{\sigma}\lambda \geq \frac{R}{1-\eta}(1-\eta)\frac{\|\varepsilon\|_2}{\sqrt{n}} \geq \underline{\sigma}\sqrt{\frac{2\log(n/\widehat{\alpha})}{n}},$$

with probability greater than $1 - \widehat{\alpha}\left(2\sqrt{2}/n + 2/\sqrt{3}\right)$.

Eventually, setting $\widehat{\alpha} = n^{-\widehat{\beta}}$ and $\widetilde{\alpha} = n^{-\widetilde{\beta}}$, we can chose the constants $\widehat{\beta} > 0, \widetilde{\beta} > 0$, so that, for $n$ large enough,

$$\frac{\widetilde{\lambda}}{\widehat{\lambda}} \leq \sqrt{2} \left( 1 - \sqrt{-2\widehat{\beta}\log n/n} \right)^{-1/2} \sqrt{\frac{\log(n/\widetilde{\alpha})}{\log(n/\widehat{\alpha})}} \leq 2\sqrt{\frac{1+\widetilde{\beta}}{1+\widehat{\beta}}} \leq \frac{C_F \wedge C_N}{2C} \ .$$

In this case, note that the probability of failure of the event $\{\widetilde{\lambda}/\widehat{\lambda} \leq (C_F \wedge C_N)/(2C)\}$ decays as a power of $n$.

**Control of $\widehat{\lambda}$.** Invoke Proposition 4 with $\eta = 1/2$ (reminding that $\lambda \geq 2R = 2\sqrt{2\log(n/\widehat{\alpha})/n}$ and that Assumption 1 holds, fulfilling (40) and (41)) and Lemma 24 (with $x = \gamma \log n$) to obtain,

$$\widehat{\lambda} = \hat{\sigma}\lambda \leq \frac{(1+\eta)\|\varepsilon\|_2\lambda}{\sqrt{n}} \leq \frac{3\sqrt{2}}{2}\sigma_0\lambda(1 + \gamma\log n/n + \sqrt{2\gamma\log n/n}) \ . \tag{31}$$

Invoke (30) to conclude the proof of Lemma 12. □

The quantity $\widehat{\lambda}$ is controlled by (31), we can conclude the proof of the theorem.

# B Proof of Theorem 2

Let us remind that by Equation (3), $\mu^0 = \sum_{j=1}^{s_0} a_j^0 \delta_{t_j^0}$. Then, let $q$ be a dual certificate of $\mu^0$ obtained by applying Lemma 13 (see Section C.1) to the set $\mathrm{supp}(\mu^0) = (t_1^0, \ldots, t_{s_0}^0)$. Recall that $q$ then interpolates the phase $v_j = a_j^0/|a_j^0|$ at the point $t_j^0$. Recall also that $q$ is a trigonometric polynomial of degree $f_c$ with $\|q\|_\infty \leq 1$. Consider $D_{\mathrm{TV}}(\hat{\mu}, \mu^0)$ the Bregman divergence of the TV-norm between the solution $\hat{\mu}$ of (4) and the target measure $\mu^0$, namely

$$D_{\mathrm{TV}}(\hat{\mu}, \mu^0) := \|\hat{\mu}\|_{\mathrm{TV}} - \|\mu^0\|_{\mathrm{TV}} - \mathfrak{R}\left(\int_{\mathbb{T}} \overline{q}(t)(\hat{\mu} - \mu^0)(\mathrm{d}t)\right) . \tag{32}$$

Since $q$ interpolates the phases of $\mu^0$, one can show that

$$\mathfrak{R}\left(\int_{\mathbb{T}} \overline{q}(t)(\hat{\mu} - \mu^0)(\mathrm{d}t)\right) = \mathfrak{R}\left(\int_{\mathbb{T}} \overline{q}(t)\hat{\mu}(\mathrm{d}t)\right) - \|\mu^0\|_{\mathrm{TV}} \leq \|\hat{\mu}\|_{\mathrm{TV}} - \|\mu^0\|_{\mathrm{TV}},$$

using Holder's inequality and $\|q\|_\infty \leq 1$. It shows that $D_{\mathrm{TV}}(\hat{\mu}, \mu^0)$ is non-negative. From this point, we consider the framework of the proof of Theorem 1. In particular, we invoke Lemma 16, Eqs (28), (30) and the control of the event $\left\{\widetilde{\lambda}/\widehat{\lambda} \leq (C_F \wedge C_N)/(2C)\right\}$ to get that there exists a constant $C > 0$ such that

$$D_{\mathrm{TV}}(\hat{\mu}, \mu^0) \leq C s_0 \widehat{\lambda} \ . \tag{33}$$

From now on, universal constants $C > 0$ may change from line to line but they do not depend on $n, \alpha, s_0, \sigma_0, \lambda, \widetilde{\lambda}$ or $\widehat{\lambda}$. Proposition 4 (with $\eta = 1/2$) and Lemma 24 (with $x = -\log\alpha$) show that

$$\widehat{\lambda} = \hat{\sigma}\lambda_0 \leq \frac{3}{\sqrt{2}}\Big(1 + \frac{\log(1/\alpha)}{n} + \sqrt{\frac{2\log(1/\alpha)}{n}}\Big)\sigma_0\lambda \ , \tag{34}$$

with probability greater than $1 - \alpha\left(\frac{2\sqrt{2}}{n} + \frac{2\sqrt{3}+6}{3}\right)$. Invoke (33) and (34) to get that

$$D_{\mathrm{TV}}(\hat{\mu}, \mu^0) \leq C\Big(1 + \frac{2\log(1/\alpha)}{n} + \sqrt{\frac{2\log(1/\alpha)}{n}}\Big)\sigma_0 s_0 \lambda \leq D_\alpha(\sigma_0, s_0, \lambda) \ , \tag{35}$$

where we define

$$D_\alpha(\sigma_0, s_0, \lambda) := C\Big(1 + \frac{\log(1/\alpha)}{n} + \sqrt{\frac{2\log(1/\alpha)}{n}}\Big)\sigma_0 s_0 \lambda \ ,$$

with $C > 0$ a universal constant that is sufficiently large to ensure the correctness of all the (forthcoming) bounds involving $D_\alpha$.

Denote $\hat{\mu} = \sum_{k=1}^{\hat{s}} \hat{a}_k \delta_{\hat{t}_k}$ a solution[4] of (4) and observe that

$$
\begin{aligned}
D_{\mathrm{TV}}(\hat{\mu}, \mu^0) &= \|\hat{\mu}\|_{\mathrm{TV}} - \|\mu^0\|_{\mathrm{TV}} - \Re\left(\int_{\mathbb{T}} \overline{q}(t)(\hat{\mu} - \mu^0)(\mathrm{d}t)\right) \ , \\
&= \|\hat{\mu}\|_{\mathrm{TV}} - \Re\left(\int_{\mathbb{T}} \overline{q}(t)\hat{\mu}(\mathrm{d}t)\right) \ , \\
&= \sum_{k=1}^{\hat{s}} \left(|\hat{a}_k| - \Re(\overline{q}(\hat{t}_k)\hat{a}_k)\right) , \\
&\geq \sum_{k=1}^{\hat{s}} |\hat{a}_k|(1 - |q(\hat{t}_k)|) \ , \\
&\geq \sum_{k=1}^{\hat{s}} |\hat{a}_k| \min\left\{(C_N/2)n^2 \min_{t \in \mathrm{supp}(\mu^0)} \mathrm{d}(t, \hat{t}_k)^2, C_F\right\} \ , \qquad (36)
\end{aligned}
$$

using Cauchy-Schwarz inequality and Lemma 13. Claims (2) and (3) follow from (35) and (36). Recall that the set of "near" points is defined as $N_j := \left\{t \in \mathbb{T};\ d(t, t_j^0) \leq \frac{c_1}{f_c}\right\}$ for some $0 < c_1 < c_0/2$, as in the papers [14, 26]; and the set of "far" points as $F := [0,1] \setminus \bigcup_{j \in [s]} N_j$. Let $q_j := q_{01,j}$ be constructed as in Section C.2 with respect to $\mathrm{supp}(\mu^0)$. In particular, Lemma 15 shows that $\|q_{01,j}\|_1 \leq \frac{C_1 s_0}{n}$ (where $s = s_0$). We get that, for all $j \in \{1, \ldots, s_0\}$,

$$
\begin{aligned}
&\Big| \sum_{\{k:\ \hat{t}_k \notin N_j\}} \hat{a}_k q_{01,j}(\hat{t}_k) + \sum_{\{k:\ \hat{t}_k \in N_j\}} |\hat{a}_k|(q_{01,j}(\hat{t}_k) - 1)\Big| \\
&\leq \sum_{\{k:\ \hat{t}_k \notin N_j\}} |\hat{a}_k||q_{01,j}(\hat{t}_k)| + \sum_{\{k:\ \hat{t}_k \in N_j\}} |\hat{a}_k||q_{01,j}(\hat{t}_k) - 1| \ , \\
&\leq \sum_{k=1}^{\hat{s}} |\hat{a}_k| \min\left\{(C_N'/2)n^2 \min_{t \in \mathrm{supp}(\mu^0)} \mathrm{d}(t, \hat{t}_k)^2, 1 - C_F\right\} \ , \\
&\leq \max\left\{\frac{1 - C_F}{C_F}, \frac{C_N'}{C_N}\right\} \times \sum_{k=1}^{\hat{s}} |\hat{a}_k| \min\left\{(C_N/2)n^2 \min_{t \in \mathrm{supp}(\mu^0)} \mathrm{d}(t, \hat{t}_k)^2, C_F\right\} \ , \\
&\leq \max\left\{\frac{1 - C_F}{C_F}, \frac{C_N'}{C_N}\right\} D_{\mathrm{TV}}(\hat{\mu}, \mu^0) \ , \\
&\leq \frac{1}{2} D_\alpha(\sigma_0, s_0, \lambda) \ . \qquad (37)
\end{aligned}
$$

using Section C.2 and (36). Furthermore, Lemma 16 shows

$$
\left|\int_{\mathbb{T}} q_{01,j}(t)(\hat{\mu} - \mu^0)(\mathrm{d}t)\right| \leq C_1 s_0 \Big(\frac{\|\mathcal{F}_n^*(\varepsilon)\|_\infty}{n} + \widehat{\lambda}\Big) \ .
$$

Using Lemma 18, with probability $1 - \alpha$, it holds that $\|\mathcal{F}_n^*(\varepsilon)\|_\infty \leq 2n\sigma_0\sqrt{\log(n/\alpha)/n}$. Invoke

[4] Recall that almost surely this solution is unique and has finite support, see Section 3.1.

Eq. (34) and recall $\lambda \geq 2\sqrt{2}\sqrt{\log(n/\alpha)}/\sqrt{n}$ to get that,

$$\left|\int_{\mathbb{T}} q_{01,j}(t)(\hat{\mu}-\mu^0)(\mathrm{d}t)\right| \leq \frac{1}{2} D_\alpha(\sigma_0, s_0, \lambda) \ , \tag{38}$$

with probability greater than $1-\alpha\left(2\sqrt{2}/n + (2\sqrt{3}+9)/3\right)$. Using inequalities (37) and (38), one can check that, for all $j \in \{1,\ldots,s_0\}$,

$$\begin{aligned}\left|a_j^0 - \sum_{\{k:\ \hat{t}_k \in N_j\}} \hat{a}_k\right| &\leq \left|\int_{\mathbb{T}} q_{01,j}(t)(\hat{\mu}-\mu^0)(\mathrm{d}t) + \sum_{\{k:\ \hat{t}_k \notin N_j\}} \hat{a}_k q_{01,j}(\hat{t}_k) + \sum_{\{k:\ \hat{t}_k \in N_j\}} \hat{a}_k (q_{01,j}(\hat{t}_k) - 1)\right| \ , \\ &\leq D_\alpha(\sigma_0, s_0, \lambda).\end{aligned}$$

This proves Claim (1).

# C Standard constructions of dual/interpolating polynomials

This section is devoted to present the different interpolating polynomials that we shall use in this paper. These polynomials are offsprings of the construction given in the pioneering paper [14] that has been recently improved by [26].

## C.1 Two constructions

**Lemma 13** (Interpolating polynomial)**.** *There exists universal positive constants $C_N$, $C'_N$ and $C_F$ such that the following holds. For any set of point $\{t_1,\ldots,t_s\}$ satisfying Assumption 2, for any $v \in \mathbb{C}^s$ such that $|v_1| \leq 1,\ldots,|v_s| \leq 1$, there exists a complex trigonometric polynomial $q_1$ of degree less than $f_c$ such that*

*(i) for all $j \in [s]$, it holds $q(t_j) = v_j$,*

*(ii) for all $j \in [s]$ and for all $t \in N_j$, it holds $|q(t)| \leq 1 - C_N \frac{n^2}{2} \mathrm{d}^2(t,t_j)$ and $|q(t) - v_j| \leq C'_N \frac{n^2}{2} \mathrm{d}^2(t,t_j)$,*

*(iii) for all $t \in F$, it holds $|q(t)| < 1 - C_F$,*

*where we recall that $n = 2f_c + 1$.*

The proof of Lemma 13 can be found using the proof of Lemma 2.2 in [25] and Lemma 2.2 in [14].

**Remark 4.** *Note that Claim (ii) leads to $|q(t)| \geq 1 - C'_N \frac{n^2}{2} \mathrm{d}^2(t,t_j)$.*

**Lemma 14** (Interpolating derivative polynomial)**.** *There exists universal positive constants $C_{N,0}$ $C_{F,0}$ such that the following holds. For any set of point $\{t_1,\ldots,t_s\}$ satisfying Assumption 2, for any $v \in \mathbb{C}^s$ such that $|v_1| \leq 1,\ldots,|v_s| \leq 1$, there exists a complex trigonometric polynomial $q_0$ of degree less than $f_c$ such that*

*(i) for all $j \in [s]$ and for all $t \in N_j$, it holds $|q_0(t) - v_j(t-t_j)| \leq C_{N,0} \frac{n}{2} \mathrm{d}^2(t,t_j)$,*

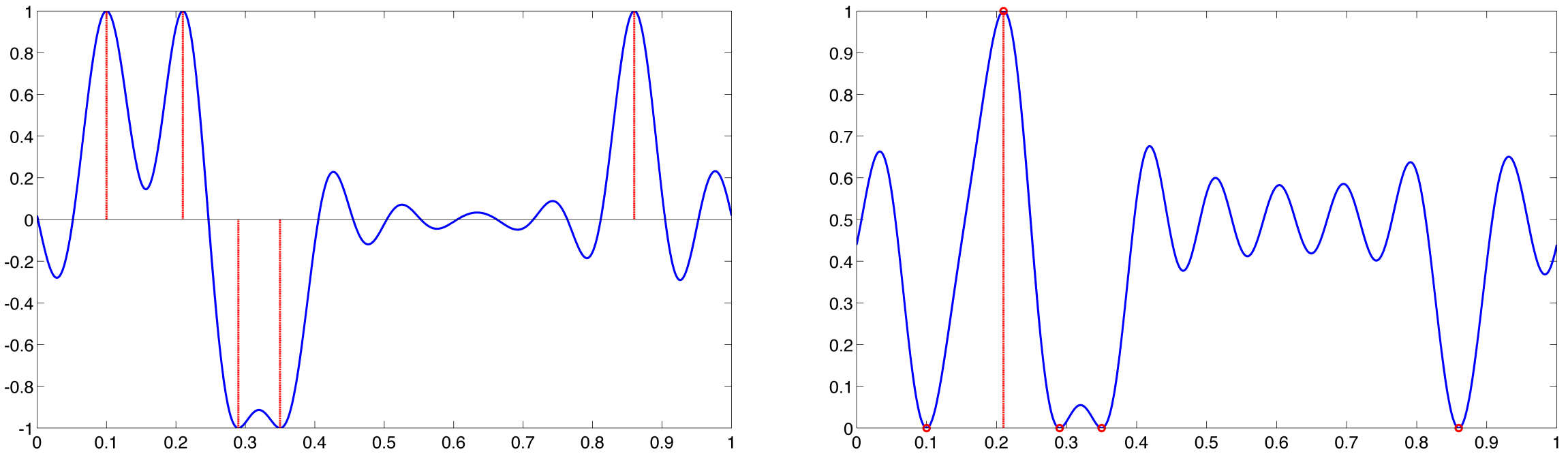

Figure 4 – Interpolating polynomial $q_1$ ($|v_1| = \cdots = |v_s| = 1$) on the left and $q_{01}$ ($|v_1| = \cdots = |v_s| = 0$ except for one $v_j = 1$) on the right.

*(ii) for all $t \in F$, it holds $|q_0(t)| < \frac{C_{F,0}}{n}$,*

*where we recall that $n = 2f_c + 1$.*

The proof of Lemma 14 can be found in the proof of Lemma 2.7 in [13] which can be improved (lowering $c_0 = 1.26$ in Assumption 2) using the paper [26].

## C.2 Dual certificates

Throughout this paper we shall use the following polynomials, see Figure 4. Consider a set of point $\{t_1, \ldots, t_s\}$ satisfying Assumption 2.

- Invoke Lemma 13 for well chosen complex numbers $|v_1| = \cdots = |v_s| = 1$ to get a ”dual certificate” that we shall denote by $q_1$.
- Fix $j \in [s]$ and invoke Lemma 13 with $v_j = 1$ and $v_i = 0$ for $i \neq j$ to define the polynomial $q_{01,j}$. In particular, the polynomial $q_{01,j}$ enjoys
  - $q_{01,j}(t_j) = 1$,
  - for all $t \in N_j$, $|1 - q_{01,j}(t)| \leq C'_N \frac{n^2}{2} \mathrm{d}^2(t, t_j)$,
  - for all $i \neq j$, for all $t \in N_i$, $|q_{01,j}(t)| \leq C'_N \frac{n^2}{2} \mathrm{d}^2(t, t_i)$,
  - for all $t \in F$, it holds $|q_{01,j}(t)| < 1 - C_F$.
- Polynomial $q_0$ is given by Lemma 14.

## C.3 Control of the Bregman divergence

**Lemma 15** ([37], Lemma 4)**.** *With the same notation as Lemma 13, there exists a universal positive constant $C_1 > 0$ such that the polynomials $q_1$, $q_{01,j}$ and $q_0$ defined in Section C.2 satisfy*

*(i)* $\|q_1\|_1 \leq \frac{C_1 s}{n}$,

*(ii)* $\|q_{01,j}\|_1 \leq \frac{C_1 s}{n}$,

*(iii)* $\|q_0\|_1 \leq \frac{C_1 s}{n^2}$,

*where we recall that* $n = 2f_c + 1$.

The proof of Lemma 15 can be found in the proof of Lemma 4 in [37].

**Lemma 16.** *With the same notation as Lemma 13, let* $\nu := \hat{\mu} - \mu^0$. *Let* $\widetilde{\alpha} \in (0,1)$ *and set* $\widetilde{\lambda} := 2\sigma_0\sqrt{\log(n/\widetilde{\alpha})/n}$. *Then, with probability greater than* $1 - \widetilde{\alpha}$, *it holds*

$$\left|\int_{\mathbb{T}} q_1(t)\nu(\mathrm{d}t)\right| \leq C_1 s\left(\widetilde{\lambda} + \widehat{\lambda}\right) \ ,$$

$$\left|\int_{\mathbb{T}} q_{01,j}(t)\nu(\mathrm{d}t)\right| \leq C_1 s\left(\widetilde{\lambda} + \widehat{\lambda}\right) \ ,$$

$$\left|\int_{\mathbb{T}} q_0(t)\nu(\mathrm{d}t)\right| \leq \frac{C_1 s}{n}\left(\widetilde{\lambda} + \widehat{\lambda}\right) \ ,$$

*where* $C_1 > 0$ *is the universal constant defined in Lemma 15 and the polynomials* $q_1$, $q_{01,j}$ *and* $q_0$ *are defined in Section C.2.*

*Proof.* We prove the first inequality, the second follows the same lines. Remind that $q_1$ is a trigonometric polynomial of degree less than $f_c$ and write

$$q_1 = \sum_{k=-f_c}^{f_c} d_k^{(1)}\phi_k \ ,$$

where we recall that $\phi_k(\cdot) = \exp(2\pi\imath k\cdot)$. Set $\Delta$ to be the following trigonometric polynomial (used also in (25))

$$\mathcal{F}_n^*(\hat{\mu} - \mu^0) = \sum_{k=-f_c}^{f_c} c_k(\hat{\mu} - \mu^0)\phi_k =: \Delta \ , \tag{39}$$

then, using Parseval's identity and Holder's inequality, we have

$$\begin{aligned}
\left|\int_{\mathbb{T}} q_1(t)\nu(\mathrm{d}t)\right| &= \left|\int_{\mathbb{T}} \sum_{k=-f_c}^{f_c} d_k^{(1)}\phi_k(t)\nu(\mathrm{d}t)\right| \ , \\
&= \left|\sum_{k=-f_c}^{f_c} d_k^{(1)}\int_{\mathbb{T}} \phi_k(t)\nu(\mathrm{d}t)\right| = \left|\sum_{k=-f_c}^{f_c} d_k^{(1)}\overline{c_k(\hat{\mu} - \mu^0)}\right|, \\
&= \left|\int_{\mathbb{T}} q_1(t)\overline{\Delta(t)}\mathrm{d}t\right| \ , \\
&\leq \|q_1\|_1\|\Delta\|_\infty \ .
\end{aligned}$$

Using Lemma 15(i) and Lemma 10, we have

$$\left|\int_{\mathbb{T}} q_1(t)\nu(\mathrm{d}t)\right| \leq C_1 s\left(\widetilde{\lambda} + \widehat{\lambda}\right) \ ,$$

as claimed. □

# D Statistical analysis

## D.1 Noise level estimation

In order to control $\hat{\sigma}$, one can use the following result given by [39].

**Proposition 17** ([39], Lemma 3.1)**.** *Suppose that for some $\eta \in (0,1)$, some $R > 0$ and some $\underline{\sigma} > 0$. Assume also that*

$$\lambda \geq \frac{R}{1-\eta} \ , \tag{40}$$

*and*

$$\lambda \frac{\|\mu^0\|_{\mathrm{TV}}}{\underline{\sigma}} \leq 2\left(\sqrt{1+(\eta/2)^2}-1\right) \ . \tag{41}$$

*Then, on the set where* $\left\{\frac{\|\mathcal{F}_n^*(\varepsilon)\|_\infty}{\sqrt{n}\|\varepsilon\|_2} \leq R\right\}$ *and* $\left\{\frac{\|\varepsilon\|_2}{\sqrt{n}} \geq \underline{\sigma}\right\}$*, one has*

$$\left|\frac{\sqrt{n}\hat{\sigma}}{\|\varepsilon\|_2} - 1\right| \leq \eta \ .$$

The proof of [39, Lemma 3.1] is elementary but non-trivial, mainly based on triangular inequalities, optimality conditions, norm convexity. It is still valid in our setting with the following notation correspondences: $\lambda \to \lambda_0$, $\lambda_{0,\varepsilon} \to R$, $\|\cdot\|_n \to \|\cdot\|_2/\sqrt{n}$, $X\beta \to \mathcal{F}_n(\mu)$, $X^T\varepsilon \to \mathcal{F}_n^*(\varepsilon)$, $\|\beta^0|_1 \to \|\mu^0\|_{TV}$.

## D.2 Control of the processes

Standard approaches in $\ell_1$-minimization are based on bounding the noise correlation $\mathcal{F}_n^*(\varepsilon)$ or its normalized version $\mathcal{F}_n^*(\varepsilon)/\|\varepsilon\|_2$ in the Concomitant Beurling Lasso case. In particular, one needs to upper bound the probabilities of the following events

$$\{\|\mathcal{F}_n^*(\varepsilon)\|_\infty \leq \lambda\} \quad \text{and} \quad \left\{\frac{\|\mathcal{F}_n^*(\varepsilon)\|_\infty}{\sqrt{n}\|\varepsilon\|_2} \leq R\right\} \ ,$$

where we recall that

$$\mathcal{F}_n^*(\varepsilon)(t) = \sum_{k=-f_c}^{f_c} \varepsilon_k \exp(2\pi\imath kt) \ ,$$

for all $t \in [0,1]$. The first event can be handled with a Rice formula for stationary Gaussian processes as in [2]. Due to the denominator, the second event cannot be described by a Gaussian process and its analysis is a bit more involved.

**Remark 5.** *A natural question could be to compare the Rice method to standard entropy arguments for computing the aforementioned events. Comparing the Rice method with entropy arguments is a well referenced discussion in the community working on the supremum of Gaussian processes, see references below. Entropy methods are indeed more general (it requires less regularity than the Rice method) but, when it comes to Gaussian processes, the Rice-Euler method may offer a competitive alternative.*

*Observe that entropy methods provide concentration inequalities with the good deviation rate but often with unknown constants in front of the exponential and a variance term to be calculated in*

*the exponent. Moreover, even in the simplest case of a Wiener process (where the expectation is known), the bounds given by standard Gaussian concentration are crude, e.g. [3, Page 61-62].*

*On the other hand, the Rice method requires regular processes and regular index sets. However, this method is sharp, see [3, Proposition 4.2] that gives an equivalent of the tail distribution. On a more general note, the Rice method gives the exact expression of the number of crossings at any level, leading to a rather sharp estimate of the tail distribution of the supremum with tractable constants. When it comes to applications, it could be interesting to have an idea of the constants appearing in the inequalities. Unfortunately, this is an issue for entropy methods (e.g. the constants in Dudley inequality are unknown).*

*In a nutshell, entropy, chaining and/or Gaussian concentration methods are very general tools that often lead to the right exponent in the rate function but their generality comes with a price when comparing it to the ground truth. A contrario, the Rice/Euler method is very specific to Gaussian processes with both regular paths and index but it often leads to better/sharp estimates of the tail distribution of the supremum of Gaussian processes.*

We start with some notation. Let

$$\begin{aligned} z^{(1)} &= (z^{(1)}_{-f_c}, \ldots, z^{(1)}_0, \ldots, z^{(1)}_{f_c}), \\ z^{(2)} &= (z^{(2)}_{-f_c}, \ldots, z^{(2)}_0, \ldots, z^{(2)}_{f_c}) \ , \end{aligned}$$

be i.i.d $\mathcal{N}_n(0, \mathrm{Id}_n)$ random vectors. Set, for any $t \in [0,1]$,

$$\begin{aligned} X(t) &= z^{(1)}_0 + \sum_{k=1}^{f_c} (z^{(1)}_k + z^{(1)}_{-k}) \cos(2\pi k t) + \sum_{k=1}^{f_c} (z^{(2)}_{-k} - z^{(2)}_k) \sin(2\pi k t)\,, \\ Y(t) &= z^{(2)}_0 + \sum_{k=1}^{f_c} (z^{(2)}_k + z^{(2)}_{-k}) \cos(2\pi k t) + \sum_{k=1}^{f_c} (z^{(1)}_k - z^{(1)}_{-k}) \sin(2\pi k t)\,, \\ Z(t) &= X(t) + \imath Y(t)\,. \end{aligned}$$

Then, note that

$$\|\sigma_0 Z\|_\infty \overset{d}{=} \|\mathcal{F}_n^*(\varepsilon)\|_\infty \quad \text{and} \quad \sup_{t\in[0,1]} \frac{|Z(t)|}{\sqrt{n}(\|z^{(1)}\|_2^2 + \|z^{(2)}\|_2^2)^{\frac12}} \overset{d}{=} \frac{\|\mathcal{F}_n^*(\varepsilon)\|_\infty}{\sqrt{n}\|\varepsilon\|_2} \ ,$$

where $\sigma_0 > 0$ is the (unknown) standard deviation of the noise $\varepsilon$.

### D.2.1 The Gaussian process

**Lemma 18.** *For a complex valued centered Gaussian random vector $\varepsilon$ as defined in (2), it holds*

$$\forall u > 0, \quad \mathbb{P}\{\|\mathcal{F}_n^*(\varepsilon)\|_\infty > u\} \le n \exp\left(-\frac{u^2}{4n\sigma_0^2}\right) \ ,$$

*where $\sigma_0 > 0$ denotes the noise level.*

*Proof.* Observe that $X(t)$ and $Y(t)$ are two independent stationary Gaussian processes with the same auto-covariance function $\Sigma$ given by

$$\forall t \in [0,1], \quad \Sigma(t) = 1 + 2\sum_{k=1}^{f_c} \cos(2\pi k t) =: \mathbf{D}_{f_c}(t) \ ,$$

where $\mathbf{D}_{f_c}$ denotes the Dirichlet kernel. Set

$$ {\sigma_n}^2 = \mathbb{V}\mathrm{ar}(X(t)) = \mathbf{D}_{f_c}(0) = n \ . \tag{42} $$

We use the following inequalities, for any $u > 0$,

$$ \mathbb{P}\{\|Z\|_\infty > u\} \le \mathbb{P}\{\|X\|_\infty > u/\sqrt{2}\} + \mathbb{P}\{\|Y\|_\infty > u/\sqrt{2}\} = 2\mathbb{P}\{\|X\|_\infty > u/\sqrt{2}\} \ , \tag{43} $$

and (by symmetry of the process $X$)

$$ \mathbb{P}\{\|X\|_\infty > u/\sqrt{2}\} \le 2\mathbb{P}\{\sup_{t\in[0,1]} X(t) > u/\sqrt{2}\} \ . \tag{44} $$

To give bounds to the right hand side of (44), we use the Rice method (see [3, page 93])

$$ \begin{aligned} \mathbb{P}\{\sup_{t\in[0,1]} X(t) > u/\sqrt{2}\} &= \mathbb{P}\{\forall t \in [0,1]; \quad X(t) > u/\sqrt{2}\} + \mathbb{P}\{U_{u/\sqrt{2}} > 0\}\,, \\ &\le \frac{1}{\sqrt{2\pi}} \int_{u/(\sqrt{2}\sigma_n)}^{+\infty} \exp\left(-\frac{v^2}{2}\right) \mathrm{d}v + \mathbb{E}(U_{u/\sqrt{2}}) \ , \end{aligned} $$

where $U_v$ is the number of up-crossings of the level $v$ by the process $X(t)$ on the interval $[0,1]$. By the Rice formula (see [3, Proposition 4.1, Page 93])

$$ \mathbb{E}(U_{u/\sqrt{2}}) = \frac{1}{2\pi}\sqrt{\mathbb{V}\mathrm{ar}(X'(t))}\frac{1}{\sigma_n}\exp\left(-\frac{u^2}{4{\sigma_n}^2}\right) \ , $$

where

$$ \mathbb{V}\mathrm{ar}(X'(t)) = -\Sigma''(0) = 2(2\pi)^2 \sum_{k=1}^{f_c} k^2 = \frac{4\pi^2}{3} f_c(f_c+1)n \ . \tag{45} $$

A Chernoff argument provides for any $w > 0$, $\int_w^{+\infty} \exp(-v^2/2)\mathrm{d}v/\sqrt{2\pi} \le \exp(-w^2/2)$, which yields

$$ \mathbb{P}\Big\{\sup_{t\in[0,1]} X(t) > \frac{u}{\sqrt{2}}\Big\} \le \exp\Big(-\frac{u^2}{4n}\Big) + \sqrt{\frac{f_c(f_c+1)}{3}}\exp\Big(-\frac{u^2}{4n}\Big) \le \frac{n}{2}\exp\Big(-\frac{u^2}{4n}\Big) \ . $$

The result follows with (44). □

#### D.2.2 The non-Gaussian process

**Lemma 19.** *It holds for all* $0 < u \le 1$,

$$ \mathbb{P}\left\{\frac{\|\mathcal{F}_n^*(\varepsilon)\|_\infty}{\sqrt{n}\|\varepsilon\|_2} > u\right\} \le \Big(2\sqrt{2} + \frac{2n}{\sqrt{3}}\Big)\Big(1 - \frac{u^2}{2}\Big)^n \ . $$

*Furthermore, it holds*

$$ \frac{\|\mathcal{F}_n^*(\varepsilon)\|_\infty}{\sqrt{n}\|\varepsilon\|_2} \le \sqrt{2} $$

*almost surely.*

*Proof.* Consider the stationary process defined for any $t \in [0,1]$ by

$$\mathcal{X}(t) := \frac{X(t)}{\sqrt{n}(\|z^{(1)}\|_2^2 + \|z^{(2)}\|_2^2)^{\frac{1}{2}}} \ .$$

Notice that Lemma 20 proves the last statement of Lemma 19. Note that the process $\mathcal{X}$ is not a Gaussian and the analysis of Section D.2.1 fails. Observe that, as in (43), it holds for any $0 < u \leq 1$

$$\mathbb{P}\left\{\frac{\|\mathcal{F}_n^*(\varepsilon)\|_\infty}{\sqrt{n}\|\varepsilon\|_2} > u\right\} \leq 2\,\mathbb{P}\left\{\sup_{t\in[0,1]} \mathcal{X}(t) > \frac{u}{\sqrt{2}}\right\} \ ,$$

and it remains to bound the right hand side term. Observe that

$$\begin{aligned}\mathbb{P}\left\{\sup_{t\in[0,1]} \mathcal{X}(t) > \frac{u}{\sqrt{2}}\right\} &= \mathbb{P}\left\{\forall t \in [0,1]; \quad \mathcal{X}(t) > \frac{u}{\sqrt{2}}\right\} + \mathbb{P}\left\{\mathcal{U}_{u/\sqrt{2}} > 0\right\} , \\ &\leq \mathbb{P}\left\{\forall t \in [0,1]; \quad \mathcal{X}(t) > \frac{u}{\sqrt{2}}\right\} + \mathbb{E}\left\{\mathcal{U}_{u/\sqrt{2}}\right\} \ ,\end{aligned}$$

where $\mathcal{U}_v$ is the number of up-crossings of the level $v$ by the process $\mathcal{X}$ on the interval $[0,1]$. Eventually, we combine Lemma 22 and Lemma 23 to get

$$\mathbb{P}\left\{\frac{\|\mathcal{F}_n^*(\varepsilon)\|_\infty}{\sqrt{n}\|\varepsilon\|_2} > u\right\} \leq \left(\frac{2\sqrt{2}}{\sqrt{2-u^2}} + \frac{2\tau_n}{\pi(2-u^2)}\right)\left(1 - \frac{u^2}{2}\right)^n \ ,$$

where $\tau_n = 2\pi\sqrt{f_c(f_c+1)}/\sqrt{3}$. The result follows. □

### D.2.3 Joint law of the process and its derivative

**Lemma 20.** *It holds*

$$(\mathcal{X}(t), \mathcal{X}'(t)) \overset{d}{=} (V_1, \tau_n V_2) \ .$$

*where $\tau_n = 2\pi\sqrt{f_c(f_c+1)}/\sqrt{3}$ and $V_1$ and $V_2$ are the first coordinates of a random vector uniformly distributed on the sphere $\mathbb{S}^{2n-1}$. For any $t \in [0,1]$, the joint density $p_{(\mathcal{X}(t),\mathcal{X}'(t))}$ of $(\mathcal{X}(t), \mathcal{X}'(t))$ is given by*

$$\forall (a,b) \in \mathbb{R}^2, \quad p_{(\mathcal{X}(t),\mathcal{X}'(t))}(a,b) = \frac{n-1}{\tau_n \pi}\Big[1 - a^2 - (b/\tau_n)^2\Big]^{n-2} \mathbb{1}_{H_n}(a,b) \ ,$$

*where $H_n := \{(a,b) \in \mathbb{R}^2; \quad a^2 + (b/\tau_n)^2 < 1\}$*

*Proof.* We start by noticing that for any $t \in [0,1]$,

$$(\mathcal{X}(t), \mathcal{X}'(t)) \overset{d}{=} \frac{1}{\sqrt{n}}(\langle V, \theta(t)\rangle, \langle V, \theta'(t)\rangle) \ , \tag{46}$$

where $V$ is uniformly distributed on the sphere $\mathbb{S}^{2n-1}$ and

$$\theta(t) = (\cos(2\pi f_c t), \cos(2\pi(f_c-1)t), \ldots, 1, \ldots, \cos(2\pi f_c t), \sin(2\pi f_c t), \ldots, 0, \ldots, -\sin(2\pi f_c t)) ,$$

$$\theta'(t) = \Big( -2\pi f_c \sin(2\pi f_c t), -2\pi(f_c-1)\sin(2\pi(f_c-1)t) \ldots, 0, \ldots, -2\pi f_c \sin(2\pi f_c t) ,$$

$$2\pi f_c \cos(2\pi f_c t), 2\pi(f_c-1)\cos(2\pi(f_c-1)t), \ldots, 0, \ldots, -2\pi f_c \cos(2\pi f_c t)\Big) .$$

The property is proved as follows. First, write

$$X(t) = \langle z, \theta(t)\rangle , \tag{47}$$

$$X'(t) = \langle z, \theta'(t)\rangle , \tag{48}$$

where

$$z = (z^{(1)}_{-f_c}, \ldots, z^{(1)}_0, \ldots, z^{(1)}_{f_c}, z^{(2)}_{-f_c}, \ldots, z^{(2)}_0, \ldots, z^{(2)}_{f_c}) \sim \mathcal{N}_{2n}(0, \mathrm{Id}_{2n}) .$$

The last term is simply obtained by derivation. Using the previous displays then $V = z/\|z\|_2$ is uniform on the sphere $\mathbb{S}^{2n-1}$, and one can check that $\mathcal{X}(t) = \langle V, \theta(t)\rangle/\sqrt{n}$ and $\mathcal{X}'(t) = \langle V, \theta'(t)\rangle/\sqrt{n}$.

Using the properties of $X$ and $X'$ given in (47) and (48), combinined with (42) and (45), for all $t \in [0,1]$ it holds that $\mathbb{V}\mathrm{ar}(X(t)) = \|\theta(t)\|_2^2 = n$ and $\mathbb{V}\mathrm{ar}(X'(t)) = \|\theta'(t)\|_2^2 = \tau_n^2 n$. Moreover, one can check that $\langle \theta(t), \theta'(t)\rangle) = 0$. Since $\theta(t)$ and $\theta'(t)$ are orthogonal and since the Haar measure on the sphere is invariant under the action of the orthogonal group, we deduce from (46) that

$$(\mathcal{X}(t), \mathcal{X}'(t)) \overset{d}{=} (V_1, \tau_n V_2) , \tag{49}$$

where $\tau_n = 2\pi\sqrt{f_c(f_c+1)}/\sqrt{3}$ and $V_1$ and $V_2$ are the first coordinates of a random vector uniformly distributed on the sphere $\mathbb{S}^{2n-1}$. A standard change of variables allows us to compute the joint law of $V_1$ and $\tau_n V_2$. Indeed, let $f$ be any continuous bounded function, using spherical coordinates one has

$$\begin{aligned}
\mathbb{E}(f(V_1, \tau_n V_2)) &= \frac{1}{S^{2n-1}} \int_{[0,\pi]^{2n-2}\times[0,2\pi)} f(\cos x_1, \tau_n \sin x_1 \cos x_2) , \\
&\qquad \sin^{2n-2} x_1 \sin^{2n-3} x_2 \cdots \sin x_{2n-2}\, \mathrm{d}x_1 \cdots \mathrm{d}x_{2n-2} \mathrm{d}x_{2n-1} , \\
&= \frac{S^{2n-3}}{S^{2n-1}} \int_{[0,\pi]^2} f(\cos x_1, \tau_n \sin x_1 \cos x_2) \sin^{2n-2} x_1 \sin^{2n-3} x_2\, \mathrm{d}x_1 \mathrm{d}x_2, \\
&= \frac{n-1}{\pi} \int_{]0,\pi[^2} f(h_n(x,y)) \sin^{2n-2} x \sin^{2n-3} y\, \mathrm{d}x \mathrm{d}y ,
\end{aligned}$$

where $S^k$ denotes the $k$-dimensional surface area of the $k$-sphere $\mathbb{S}^k \subset \mathbb{R}^{k+1}$ and $h_n$ is defined from $]0,\pi[^2$ onto $H_n := \{(a,b) \in \mathbb{R}^2;\quad a^2 + (b/\tau_n)^2 < 1\}$ by $h_n(x,y) = (\cos x, \tau_n \sin x \cos y)$. Observe that $h_n$ is a $\mathcal{C}^1$-diffeomorphism whose Jacobian determinant at point $(x,y)$ is $\tau_n \sin^2 x \sin y$ and its inverse function is $h_n^{-1}(a,b) = (\arccos a, \arccos(b/(\tau_n\sqrt{1-a^2})))$. By the change of variables given by $h_n$, it holds

$$\begin{aligned}
\mathbb{E}(f(V_1, \tau_n V_2)) &= \frac{n-1}{\pi} \int_{H_n} f(a,b) \frac{1}{\tau_n} \sin^{2n-4}(\arccos a) \sin^{2n-4}(\arccos(b/(\tau_n\sqrt{1-a^2})))\, \mathrm{d}a\mathrm{d}b , \\
&= \frac{n-1}{\tau_n \pi} \int_{H_n} f(a,b) \Big[(1-a^2)^{n-2}\big(1 - \frac{b^2}{\tau_n^2(1-a^2)}\big)^{n-2}\Big]\, \mathrm{d}a\mathrm{d}b ,
\end{aligned}$$

using that $\sin(\arccos(t)) = \sqrt{1-t^2}$. From (49), we deduce that $(\mathcal{X}(t), \mathcal{X}'(t))$ has a density $p_{(\mathcal{X}(t),\mathcal{X}'(t))}$ with respect to the Lebesgue measure and it holds

$$p_{(\mathcal{X}(t),\mathcal{X}'(t))}(a,b) = \frac{n-1}{\tau_n \pi}(1-a^2)^{n-2}\Big[1 - \frac{b^2}{\tau_n^2(1-a^2)}\Big]^{n-2} \mathbb{1}_{H_n}(a,b) \ ,$$

for all $(a,b) \in \mathbb{R}^2$. □

We derive the following useful description of the law of $\mathcal{X}(t)$.

**Lemma 21.** *It holds $\mathcal{X}(t)$ has the same law as the first coordinate $V_1$ of a random vector uniformly distributed on the sphere $\mathbb{S}^{2n-1}$. For any $t \in [0,1]$, the density $p_{\mathcal{X}(t)}$ of $\mathcal{X}(t)$ is given by*

$$\forall a \in \mathbb{R}, \quad p_{\mathcal{X}(t)}(a) = \frac{\Gamma(n)}{\sqrt{\pi}\Gamma(n-1/2)}\left(1-a^2\right)^{n-3/2} \mathbb{1}_{[-1;1]}(a) \ ,$$

*with $\Gamma$ the Gamma function.*

*Proof.* Let $g$ be any continuous bounded function, using spherical coordinates one has

$$\begin{aligned}
\mathbb{E}(g(V_1)) &= \frac{1}{S^{2n-1}} \int_{[0,\pi]^{2n-2}\times[0,2\pi)} g(\cos x_1) \\
&\qquad \sin^{2n-2} x_1 \sin^{2n-3} x_2 \cdots \sin x_{2n-2}\, \mathrm{d}x_1 \cdots \mathrm{d}x_{2n-2}\mathrm{d}x_{2n-1} \ , \\
&= \frac{S^{2n-2}}{S^{2n-1}} \int_{[0,\pi]} g(\cos x_1) \sin^{2n-2} x_1 \, \mathrm{d}x_1 \ , \\
&= \frac{\Gamma(n)}{\sqrt{\pi}\Gamma(n-1/2)} \int_{]0,\pi[} g(\cos x_1) \sin^{2n-2} x_1 \, \mathrm{d}x_1 \ .
\end{aligned}$$

Using the change of variable $a = \cos x_1$, one gets

$$\begin{aligned}
\mathbb{E}(g(V_1)) &= \frac{\Gamma(n)}{\sqrt{\pi}\Gamma(n-1/2)} \int_{-1}^{1} g(a) \sin^{2n-3}(\arccos a)\, \mathrm{d}a \ , \\
&= \frac{\Gamma(n)}{\sqrt{\pi}\Gamma(n-1/2)} \int_{-1}^{1} g(a) \left(\sqrt{1-a^2}\right)^{2n-3} \mathrm{d}a \ ,
\end{aligned}$$

which ends the proof. □

#### D.2.4 Trajectories uniformly above a level

**Lemma 22.** *For all $0 < u < \sqrt{2}$, it holds*

$$\mathbb{P}\Big\{\forall t \in [0,1]; \quad \mathcal{X}(t) > \frac{u}{\sqrt{2}}\Big\} \leq 2\Big(1 - \frac{u^2}{2}\Big)^{n-\frac{1}{2}} \ .$$

*Proof.* We simply consider the elementary bound

$$\mathbb{P}\Big\{\forall t \in [0,1]; \quad \mathcal{X}(t) > \frac{u}{\sqrt{2}}\Big\} \leq \mathbb{P}\Big\{\mathcal{X}(0) > \frac{u}{\sqrt{2}}\Big\} \ ,$$

where we recall that

$$\mathcal{X}(0) = \frac{\sum_{k=-f_c}^{f_c} z_k^{(1)}}{\sqrt{n}(\|z^{(1)}\|_2^2 + \|z^{(2)}\|_2^2)^{\frac{1}{2}}} \overset{d}{=} \frac{z_0^{(1)}}{(\|z^{(1)}\|_2^2 + \|z^{(2)}\|_2^2)^{\frac{1}{2}}} \ ,$$

where the last equality in distribution holds thanks to (49). Then, we make use of the following inequalities as in [39]

$$\begin{aligned}
\mathbb{P}\Big\{\mathcal{X}(0) > \frac{u}{\sqrt{2}}\Big\} &= \mathbb{P}\Big\{\frac{(z_0^{(1)})^2}{\|z^{(1)}\|_2^2 + \|z^{(2)}\|_2^2} > \frac{u^2}{2}\Big\} \ , \\
&= \mathbb{P}\left\{\left(1 - \frac{u^2}{2}\right)(z_0^{(1)})^2 > \frac{u^2}{2}\Big(\sum_{k=1}^{f_c}(z_k^{(1)})^2 + (z_{-k}^{(1)})^2 + \|z^{(2)}\|_2^2\Big)\right\} \ , \\
&= \int_0^{+\infty} \mathbb{P}\Big\{(z_0^{(1)})^2 > \frac{tu^2}{2-u^2}\Big\} f_{\chi^2_{2n-1}}(t)\mathrm{d}t \ , \\
&\leq \int_0^{+\infty} 2\exp\Big(-\frac{tu^2}{4-2u^2}\Big) f_{\chi^2_{2n-1}}(t)\mathrm{d}t \ , \\
&= 2\mathbb{E}\left[\exp\left(-\frac{u^2 Z}{4-2u^2}\right)\right] \ , \\
&= 2\left(\frac{1}{1+\frac{u^2}{2-u^2}}\right)^{\frac{2n-1}{2}} \ , \\
&= 2\left(1 - \frac{u^2}{2}\right)^{n-\frac{1}{2}} \ .
\end{aligned}$$

where $f_{\chi^2_{2n-1}}$ denotes the density function of the chi-squared distribution with $2n-1$ degrees of freedom and $Z$ is distributed with respect to this distribution. Note that we have used Fubini's theorem and a Chernoff argument providing for any $v > 0$, $1 - \Psi(v) \leq \exp(-v^2/2)$ where $\Psi$ denotes the cumulative distribution function of the standard normal distribution. □

#### D.2.5 Number of up-crossings

**Lemma 23.** *It holds*

$$\mathbb{E}\{\mathcal{U}_{u/\sqrt{2}}\} \leq \frac{\tau_n}{2\pi}\Big(1 - \frac{u^2}{2}\Big)^{n-1} \ ,$$

*where we recall that* $\tau_n = 2\pi\sqrt{f_c(f_c+1)}/\sqrt{3}$.

*Proof.* First observe that the joint density $p_{(\mathcal{X}(t),\mathcal{X}'(t))}$ of $(\mathcal{X}(t), \mathcal{X}'(t))$ is a compactly supported continuous function that does not depend on $t$ by Lemma 20. In order to bound $\mathbb{E}(\mathcal{U}_{u/\sqrt{2}})$, we make use of the following result described in [3, Page 79]

$$\mathbb{E}\Big\{\mathcal{U}_{u/\sqrt{2}}\Big\} \leq \int_0^1 \mathrm{d}t \int_0^\infty x p_{(\mathcal{X}(t),\mathcal{X}'(t))}(u/\sqrt{2}, x)\mathrm{d}x \ . \tag{50}$$

Lemma 20 and an elementary change of variables (namely $z = (x/\tau_n)^2$ and $y = 1 - u^2/2 - z$) give

$$\mathbb{E}\Big\{\mathcal{U}_{u/\sqrt{2}}\Big\} \leq \frac{n-1}{\tau_n \pi} \int_0^{\tau_n\sqrt{1-u^2/2}} x\Big(1 - \frac{u^2}{2} - \frac{x^2}{\tau_n^2}\Big)^{n-2} \mathrm{d}x \ ,$$
$$= \frac{(n-1)\tau_n}{2\pi} \int_0^{1-u^2/2} x^{n-2} \mathrm{d}x \ ,$$
$$= \frac{\tau_n}{2\pi}\Big(1 - \frac{u^2}{2}\Big)^{n-1} \ ,$$

as claimed. □

### D.3 Concentration

#### D.3.1 The chi-squared distribution

We control the Chi-square deviation using a standard lemma recalled here.

**Lemma 24.** *[31, Lemma 1] If $\varepsilon$ is a complex valued centered Gaussian random variable defined by $\varepsilon \overset{d}{=} \varepsilon^{(1)} + \imath\varepsilon^{(2)}$ where the real part $\varepsilon^{(1)} = \Re(\varepsilon)$ and the imaginary part $\varepsilon^{(2)} = \Im(\varepsilon)$ are i.i.d. random vectors $\mathcal{N}_n(0, \sigma_0^2 \operatorname{Id}_n)$. It holds*

$$\mathbb{P}\left\{\|\varepsilon\|_2^2 \leq 2n\sigma_0^2\left(1 - \sqrt{\frac{2x}{n}}\right)\right\} \leq \exp(-x) \ ,$$
$$\mathbb{P}\left\{\|\varepsilon\|_2^2 \geq 2n\sigma_0^2\left(1 + \frac{x}{n} + \sqrt{\frac{2x}{n}}\right)\right\} \leq \exp(-x) \ .$$

*Proof.* Take $D = 2n$ and $a_i = \sigma_0^2$ in [31, Lemma 1]. □

## E Convexity tools

We first remind some classical results from standard convex analysis, see [4, Example 13.8, Proposition 13.20, and Example 13.6].

### E.1 Convexity reminder

The sub-gradient of a convex function $f : \mathbb{C}^d \to \mathbb{R}$ at $x$ is defined as

$$\partial f(x) = \{z \in \mathbb{C}^d; \forall y \in \mathbb{R}^d, f(x) - f(y) \geq \langle z, (x - y)\rangle\} \ . \tag{51}$$

We denote $f^*$ the Fenchel-conjugate of $f$, $f^*(z) = \sup_{w \in \mathbb{C}^d} \langle w, z\rangle - f(w)$. In this section, for convexity analysis purpose, we will denote by $I_C$ the indicator function of a set $C$ defined as

$$I_C : \mathbb{C}^d \to \mathbb{R}, \quad I_C(x) = \begin{cases} 0, & \text{if } x \in C \ , \\ +\infty, & \text{otherwise} \ . \end{cases} \tag{52}$$

**Lemma 25.** *For a convex function $f : \mathbb{C}^n \to \mathbb{R}$, its perspective function is the function*

$$\operatorname{persp}(f) : \mathbb{C}^n \times \mathbb{R} \to \mathbb{R}, \quad (x, t) \mapsto \begin{cases} tf(\frac{x}{t}), & \text{if } t > 0 \ , \\ +\infty, & \text{otherwise} \ . \end{cases} \tag{53}$$

*Its Fenchel-conjugate* $\operatorname{persp}(f)^*$ *reads* $\operatorname{persp}(f)^* = I_{\{(z,u)\in\mathbb{C}^n\times\mathbb{R}\,:\, u+f^*(z)\leq 0\}}$.

**Lemma 26.** *For a function $f : \mathbb{C}^n \to \mathbb{R}$ and for any $z \in \mathbb{C}^n$ one has the following properties for the Fenchel-conjugate:*

- $(\tau_z(f))^* = f^* + \langle\cdot, z\rangle$, *where* $\tau_z(f) = f(\cdot - z)$.
- $(f + \langle\cdot, z\rangle)^* = \tau_z(f^*)$.
- *For* $(\|\cdot\|^2/2)^* = \|\cdot\|^2/2$.

## E.2 Proof of Proposition 5

Applying Lemma 25 and Lemma 26 to the function $f = \|\cdot\|^2/2$ and $\operatorname{persp}(f)(x,t) = \|x\|_2^2/(2t)$, provides $\operatorname{persp}(f)^* = I_{\{(z,u)\in\mathbb{C}^n\times\mathbb{R}\,:\, u+\|z\|^2/2\leq 0\}}$.

We can now dualize Problem (4). First remark that the objective function in (4) can be written as $P_\lambda(\mu,\sigma) = \operatorname{persp}(\|\cdot\|^2/2)(\mathcal{F}_n(\mu) - y, n\sigma) + \sigma/2 + \lambda\|\mu\|_{\mathrm{TV}}$. For $h : \mathbb{C}^n \times \mathbb{R} \to \mathbb{R}$ defined by $h = \operatorname{persp}(\|\cdot\|^2/2)$, one can write the primal in the form $P_\lambda(\mu,\sigma) = \tau_z(h)(\mathcal{F}_n(\mu), n\sigma) + \sigma/2 + f(\mu,\sigma)$, where $z = (y, 0) \in \mathbb{C}^n \times \mathbb{R}$ and $f(\mu,\sigma) = \lambda\|\mu\|_{\mathrm{TV}} + I_{\mathbb{R}_{++}}$. Then, we can apply [4, Proposition, 19.18], with $g : \mathbb{C}^n \times \mathbb{R} \to \mathbb{R}$ by $g(\cdot,\sigma) = \tau_z(h)(\cdot, n\sigma) + \sigma/2$ and $L = \mathcal{F}_n$. This leads to the Lagrangian formulation

$$\mathcal{L}(\mu, \sigma, c, t) := f(\mu,\sigma) + \langle\mathcal{F}_n(\mu), c\rangle + \sigma t - g^*(c,t) \ .$$

where $g^*(c,t) = \langle c, y\rangle + I_C(c,t)$ for $C = \{(z,u) \in \mathbb{C}^n \times \mathbb{R} : u + n\|z\|^2/2 \leq 1/2\}$

Since strong duality holds, the primal problem is equivalent to finding a saddle point of the Lagrangian. Any such saddle point $(\hat{\mu}, \hat{\sigma}, \hat{c}, \hat{t})$ satisfies on the one hand $\|\mathcal{F}_n^*\hat{c}\|_\infty \leq \lambda$, and on the other hand $\hat{c} = (\mathcal{F}_n(\hat{\mu}) - y)/(n\hat{\sigma})$ and $\hat{t} = 1/2 - \|\mathcal{F}_n(\hat{\mu}) - y\|^2/(2n(\hat{\sigma})^2) = 1/2 - n\|\hat{c}\|^2/2$. The dual problem can also be obtained from the aforementioned theorem:

$$\min_{(c,t)\in\mathbb{C}^n\times\mathbb{R}} f^*(\mathcal{F}_n^*(c), t) + g^*(-c, -t) \ , \tag{54}$$

where

$$f^*(\mathcal{F}_n^*(c), t) = \begin{cases} 0, & \text{if } \|\mathcal{F}_n^* c\|_\infty \leq \lambda \text{ and } t \leq 0 \ . \\ +\infty, & \text{otherwise} \ . \end{cases} \tag{55}$$

Hence the dual problem reads

$$\begin{aligned} &\min_{(c,t)\in\mathbb{C}^n\times\mathbb{R}} \langle y, -c\rangle \ , \\ &\text{s.t. } \|\mathcal{F}_n^*(c)\|_\infty \leq \lambda, t \leq 0 \text{ and } -t + n\|c\|^2/2 \leq 1/2 \ . \end{aligned} \tag{56}$$

Re-parameterizing the dual by performing $c \leftarrow c/\lambda$ and taking $t \leftarrow -t$ leads to:

$$\begin{aligned} &\min_{(c,t)\in\mathbb{C}^n\times\mathbb{R}} \langle -y, \lambda c\rangle \ , \\ &\text{s.t.} \|\mathcal{F}_n^*(c)\|_\infty \leq 1, t \geq 0 \text{ and } t + n\lambda^2\|c\|^2/2 \leq 1/2 \ . \end{aligned} \tag{57}$$

Finally, the dual problem of (4) reads

$$\hat{c} \in \operatorname*{arg\,min}_{c \in \tilde{\mathcal{D}}_n} \langle -y, \lambda c \rangle \quad , \tag{58}$$

where $\tilde{\mathcal{D}}_n = \left\{(c,u) \in \mathbb{C}^n \times \mathbb{R} : \|\mathcal{F}_n^* c\|_\infty \le 1, u \ge 0, n\lambda^2 \|c\|^2/2 + u \le 1/2\right\}$. However, one can easily show that Problem (10) is equivalent to (58). Indeed, one can write that $\tilde{\mathcal{D}}_n = \cup_{u \ge 0} \mathcal{D}_{n,u}$ where $\mathcal{D}_{n,u} = \left\{c \in \mathbb{C}^n : \|\mathcal{F}_n^*(c)\|_\infty \le 1, n\lambda^2\|c\|^2/2 + u \le 1/2\right\}$. Moreover, one can remark that for all $u \ge 0$, $\mathcal{D}_{n,u} \subset \mathcal{D}_{n,0}$ and then for all $u \ge 0$,

$$\min_{c \in \mathcal{D}_{n,u}} \langle -y, \lambda c \rangle \ge \min_{c \in \mathcal{D}_{n,0}} \langle -y, \lambda c \rangle \quad .$$

With the previous remark, one can infer that

$$\min_{c \in \tilde{\mathcal{D}}_n} \langle -y, \lambda c \rangle = \min_{c \in \mathcal{D}_{n,0}} \langle -y, \lambda c \rangle \,,$$

the conclusion follows by noting that $\mathcal{D}_{n,0} = \mathcal{D}_n$. Finally, the dual problem can be written as follows

$$\hat{c} \in \operatorname*{arg\,max}_{c \in \mathcal{D}_n} \langle y, \lambda c \rangle \quad .$$

Equation (11) is the consequence of considering the Lagrangian formulation of the following constrained problem:

$$\begin{aligned} (z, \hat{\mu}, \hat{\sigma}) \in \operatorname*{arg\,min}_{(z,\mu,\sigma) \in E^* \times \mathbb{R}_{++}} & \frac{1}{2n\sigma}\|z\|_2^2 + \frac{\sigma}{2} + \lambda \|\mu\|_{\mathrm{TV}}, \\ \text{s.t. } & z = y - \mathcal{F}_n(\mu). \end{aligned} \tag{59}$$

## E.3 Proof of Proposition 6

We present the proof of Proposition 6 here.

(i) $\Rightarrow$ (ii): Let us choose $\lambda \in ]0, \lambda_{\min}(y)]$, then $n\lambda^2 \|\hat{c}^{(\mathrm{BME})}\|_2 \le n(\lambda_{\min}(y))^2 \|\hat{c}^{(\mathrm{BME})}\|_2 \le 1$. Hence, $\hat{c}^{(\mathrm{BME})} \in \mathcal{D}_n$, and since $\mathcal{D}_n \subset \{c \in \mathbb{C}^n : \|\mathcal{F}_n^*(c)\|_\infty \le 1\}$, then $\hat{c} = \hat{c}^{(\mathrm{BME})}$.

(ii) $\Rightarrow$ (iii): Assume that $\hat{c} = \hat{c}^{(\mathrm{BME})}$, then $y = n\hat{\lambda}\hat{c}^{(\mathrm{BME})} + \mathcal{F}_n(\hat{\mu})$ thanks to Eq. (11) and $\mathcal{F}_n(\hat{\mu}^{(\mathrm{BME})}) = y$ thanks to Eq. (17). Moreover, one has $\langle y, \hat{c}^{(\mathrm{BME})} \rangle = \|\hat{\mu}\|_{\mathrm{TV}}$ and it holds that $\lambda\langle y, \hat{c}\rangle = \|y - \mathcal{F}_n(\hat{\mu})\|^2/2n\hat{\sigma} + \hat{\sigma}/2 + \lambda\|\hat{\mu}\|_{\mathrm{TV}}$ by strong duality. The only way the last equation holds is when $\hat{\sigma} = 0$ and that $y = \mathcal{F}_n(\hat{\mu})$.

(iii) $\Rightarrow$ (i): Assume that $\hat{\sigma} = 0$, this leads to $\hat{\lambda} = 0$ thanks to the definition of $\hat{\lambda}$ below (11). Thanks to Eq. (11), $y = \mathcal{F}_n(\hat{\mu})$. This means that $(\hat{\mu}, \hat{\sigma})$ is solution of the problem

$$(\hat{\mu}, \hat{\sigma}) \in \operatorname*{arg\,min}_{\substack{(\mu,\sigma) \in E^* \times \mathbb{R}_{++} \\ y = \mathcal{F}_n(\mu)}} \frac{1}{2n\sigma}\|y - \mathcal{F}_n(\mu)\|_2^2 + \frac{\sigma}{2} + \lambda\|\mu\|_{\mathrm{TV}} \quad . \tag{60}$$

and so

$$\hat{\mu} \in \operatorname*{arg\,min}_{\substack{\mu \in E^* \\ y = \mathcal{F}_n(\mu)}} \lambda \|\mu\|_{\mathrm{TV}} \quad . \tag{61}$$

*i.e.*, $\hat{\mu} = \hat{\mu}^{(\mathrm{BME})}$.

By strong duality in Problem (4), one has $\lambda\|\hat{\mu}\|_{\mathrm{TV}} = \lambda\langle \hat{c}, y\rangle$ and by strong duality in Problem (18), $\lambda\|\hat{\mu}^{(\mathrm{BME})}\|_{\mathrm{TV}} = \lambda\langle \hat{c}^{(\mathrm{BME})}, y\rangle$. Hence $\langle \hat{c}, y\rangle = \langle \hat{c}^{(\mathrm{BME})}, y\rangle$ and one can choose $\hat{c}^{(\mathrm{BME})}$ as a dual optimal solution for Problem (10). So $\|\hat{c}^{(\mathrm{BME})}\|_2^2 \leq 1/(n\lambda^2)$, and (i) holds by definition of $\lambda_{\min}$.

We now prove the last statement of the proposition. Since $\|\hat{p}\|_\infty \leq 1$, Parseval's inequality leads to $\|\hat{c}\|_2 \leq 1$. If $\lambda < 1/\sqrt{n}$ then $\lambda^2 n\|\hat{c}\|_2 \leq \lambda^2 n < 1$, this means that the $\ell_2$ constraint in the dual formulation (10) is not saturated. Then $\hat{c} = \hat{c}^{(\mathrm{BME})}$ and using (ii) $\Leftrightarrow$ (i), we deduce that $\lambda_{\min}(y) \geq 1/\sqrt{n}$.

## E.4 Proof of Proposition 7

First note that if $\lambda \leq \lambda_{\min}(y)$, by Proposition 6 there is overfitting which contradicts the assumption made in Section 2.1. Secondly, if $\lambda > \lambda_{\max}(y)$, then by Remark 3, $\hat{\mu} = 0$ a scenario we are not interested in. Now, with Eqs (11) and (12), one can check that $\hat{c} = y/(n\widehat{\lambda}) = y/(\sqrt{n}\lambda\|y\|)$. Since $y$ is a Gaussian vector, $\hat{p} = \mathcal{F}_n^*(\hat{c})$ almost surely has a non-constant modulus.

Set that $\lambda \in [\lambda_{\min}(y); \lambda_{\max}(y)]$. Let us suppose that the polynomial $|\hat{p}|^2$ is of constant modulus, then it can be written as $p = v\varphi_k$ with $v \in \mathbb{C}$ and $\varphi_k(\cdot) = \exp(2\pi\imath k\cdot)$ for some $k \in [\![-f_c, f_c]\!]$. Note that if $|v| < 1$, using Holder's inequality on (15) leads to $\hat{\mu} = 0$. Now if $|v| = 1$, we also have $\hat{c} \in \mathcal{D}_n$, in particular $\|\hat{c}\|_2 \leq 1/(\sqrt{n}\lambda)$, leading to $|v| \leq 1/(\sqrt{n}\lambda)$. However, since $\lambda_{\min}(y) > 1/\sqrt{n}$, it turns out that $|v| < 1$, which contradicts $|v| = 1$. One can then conclude that a dual polynomial of constant modulus never occurs in the CBLasso setup, provided that $\lambda \in [\lambda_{\min}(y); \lambda_{\max}(y)]$.